\documentclass[
 reprint,
 amsmath,amssymb,
 aps,
prb,
]{revtex4-2}

\usepackage{graphicx}
\usepackage{dcolumn}
\usepackage[colorlinks=true,citecolor=blue]{hyperref}
\usepackage{xcolor}
\usepackage{bm}
\usepackage{tabularx}
\usepackage[table]{xcolor}
\usepackage{comment}

\begin{document}

\preprint{APS/123-QED}

\title{Complex field-induced magnetic phases and anisotropic magnetotransport in off-stoichiometric CeCuBi$_2$} 

\author{Vikas Chahar$^{\dagger,1}$}
\author{Shivam Rathod$^{\dagger,1}$}
\author{Sneh$^1$}
\author{Lipika$^1$}
\author{Shobha Singh$^1$}
\author{Balwant Singh Chauhan$^1$}
\author{Subhadeep Bej$^1$}
\author{Weida Yin$^2$}
\author{Yin-Chen Huang$^2$}
\author{Rie Y. Umetsu$^{2,3}$}
\author{Ratnamala Chatterjee$^1$}
\author{Kaustuv Manna$^1$}
\email{kaustuvmanna@iitd.ac.in\\  \textcolor{black}{$\dagger$ These authors contributed equally to this work.}}

\affiliation{$^1$Department of Physics, Indian Institute of Technology Delhi, New Delhi 110016,~India\\ 
$^2$Institute for Materials Research, Tohoku University, Sendai 980-8577,~Japan\\
$^3$Center for Science and Innovation in Spintronics, Tohoku University, Sendai 980-8577, Japan}





\begin{abstract}

We report a detailed study on the structural, angle-dependent magnetic and magnetotransport properties of highly anisotropic off-stoichiometric CeCuBi$_2$ single crystals. Our results reveal CeCuBi$_2$ as an anisotropic Kondo antiferromagnet exhibiting complex field-induced magnetic behavior and unusual magnetotransport properties. Magnetic susceptibility and specific heat measurements reveal antiferromagnetic (AFM) ordering below $T_N\sim14$ K with strong anisotropy and weak heavy-fermion behavior. Electrical transport measurements show highly anisotropic resistivity and a broad hump around $\sim47$~K, indicative of Kondo-driven heavy-fermion behavior. Magnetization measurements reveal multiple field-induced metamagnetic phases, while AC susceptibility measurements indicate slow spin dynamics and spin-glass-like behavior in intermediate field-induced magnetic states. Furthermore, we observe large and strongly anisotropic magnetotransport responses, including room-temperature magnetoresistance of $\sim22\%$ at 300 K and 9~T and butterfly-like anisotropic magnetoresistance with AMR values reaching $\sim10.9\%$. These results highlight a strong interplay among Kondo correlations, magnetic anisotropy, and field-tunable spin configurations, making CeCuBi$_2$ a possible platform for exploring correlated and anisotropic quantum phenomena.

\end{abstract}


\maketitle


\section{\label{sec:level1}INTRODUCTION}

Recently cerium-based Ce$T$$X$$_2$ intermetallic compounds ($T$ = Ag, Ni, Au, Cu and $X$ = Sb, Bi) have attracted significant interest because of their rich magnetic and electronic properties arising from strong electronic correlations~\cite{ye1996novel,adriano2014physical,thamizhavel2003low,nicklas2001response,piva2018high,hossain1999antiferromagnetic}. These materials crystallize in a layered tetragonal structure with $P4/nmm$ symmetry and provide an excellent platform to study the interplay between crystalline electric field (CEF) effects, Ruderman–Kittel–Kasuya–Yosida (RKKY) exchange interactions, and hybridization between localized Ce 4$f$ electrons and conduction electrons associated with the Kondo effect. The competition between the RKKY interaction, with $T{\mathrm{_{RKKY}}} \sim [J N(E_F)]^2$, and the Kondo interaction, with $T_K \sim \exp[-1/JN(E_F)]$, governs the magnetic and electronic ground states of these heavy-fermion systems~\cite{seo2012pressure}. Here, $J$ denotes the exchange coupling between Ce 4$f$ moments and conduction electrons, and $N(E_F)$ is the density of states at the Fermi level. Since both interactions depend on $J N(E_F)$, external parameters such as magnetic field, pressure, and chemical substitution can effectively tune the magnetic ordering and correlated electronic states.

The interplay among these competing interactions gives rise to a variety of emergent quantum phenomena, including complex ferro- and antiferromagnetic ordering, heavy-fermion behavior, non-Fermi-liquid states, unconventional superconductivity, and strong magnetic anisotropy. For instance, CeCu$_2$Si$_2$ and CeNi$_{0.8}$Bi$_2$ show unconventional heavy-fermion superconductivity~\cite{steglich1979superconductivity,Mizoguchi2011}, while several members of the Ce$T$$X$$_2$ family show intricate magnetic ground states and field-induced magnetic phase transitions~\cite{ye1996novel,adriano2014physical,thamizhavel2003low,nicklas2001response,piva2018high,hossain1999antiferromagnetic}.

Among the Ce$T$$X$$_2$ family, CeCuBi$_2$ is particularly intriguing because it combines strong magnetic anisotropy with tunable field-induced electronic and magnetic states. CeCuBi$_2$ behaves as an Ising-type antiferromagnet with Néel temperature $T_N \sim 16$ K~\cite{jesus2014evolution}. X-ray magnetic diffraction studies performed below $T_N$ revealed a commensurate antiferromagnetic structure with propagation vector $\left(\frac{1}{2},\frac{1}{2},0\right)$ and Ce magnetic moments aligned along the easy $c$-axis~\cite{adriano2014physical}. The quasi-two-dimensional layered crystal structure together with strong hybridization between localized Ce 4$f$ electrons and conduction electrons gives rise to pronounced magnetic and electronic anisotropy in this compound.

\begin{figure*}[t]
  \centering

  \includegraphics[width=\textwidth]{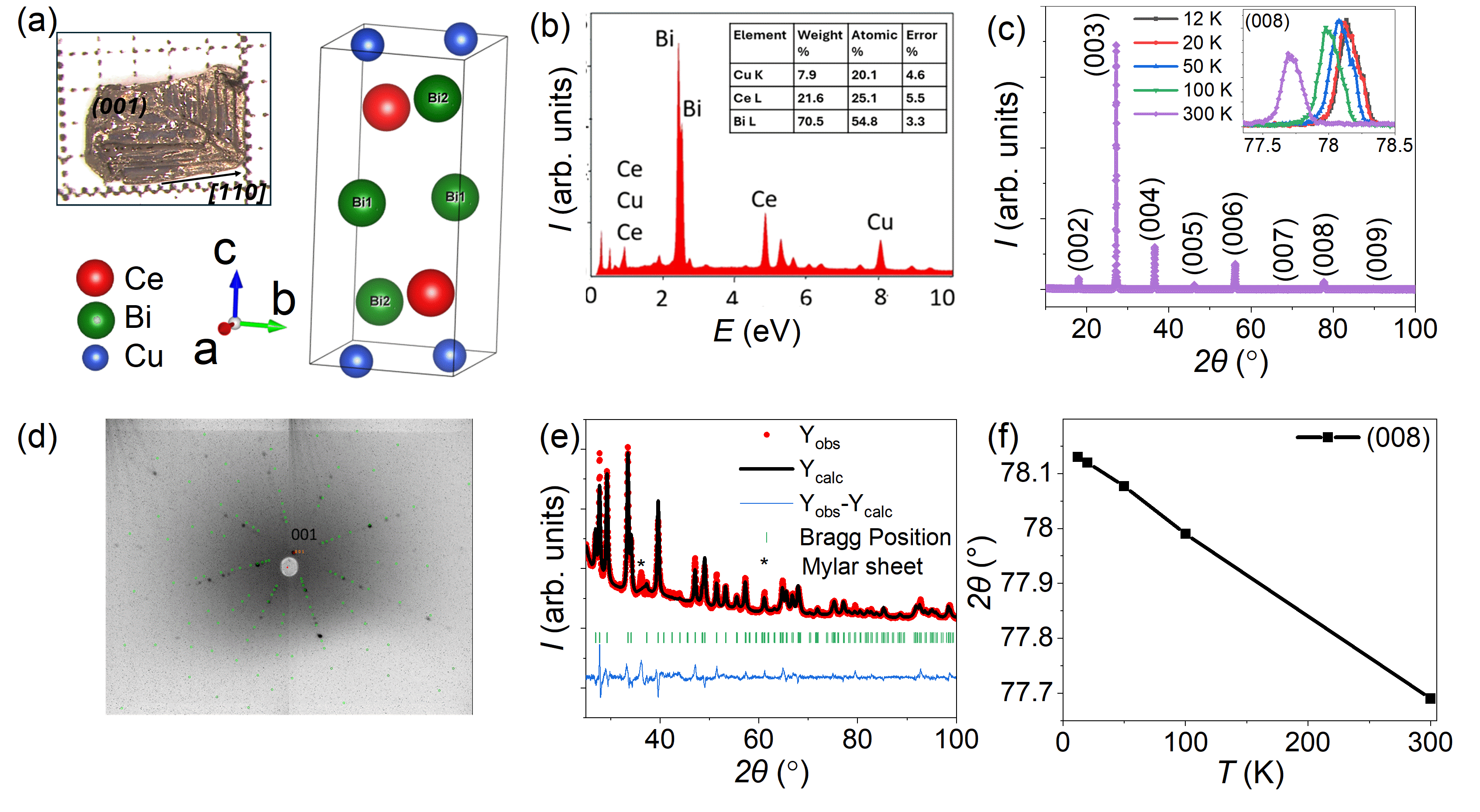}%

  \caption{\label{fig:one}
  (a) Optical image of the $\mathrm{CeCuBi_2^{os}}$ single crystal, tetragonal unit-cell structure of $\mathrm{CeCuBi_2^{os}}$, where Bi occupy two lattice sites Bi1 and Bi2. (b) EDS spectra collected at one region of the crystal. (c) XRD pattern of $\mathrm{CeCuBi_2^{os}}$ crystal surface indexed with (00l) tetragonal basal planes. Inset shows an enlarged view of the (008) XRD peak of the $\mathrm{CeCuBi_2^{os}}$ single crystal measured at different temperatures.  (d) Refined Laue diffraction with (001) plane of the crystal. \textcolor{black}{(e)} Rietveld refinement of the powder X-ray diffraction data of $\mathrm{CeCuBi_2^{os}}$. Peaks marked with (*) in the XRD pattern arise from the mylar sheet used to prevent powder oxidation. The red curve represents the observed intensity, the black curve corresponds to the calculated intensity, the green tick marks indicate the allowed Bragg reflection positions (2$\theta$), and the blue curve shows the difference between the observed and calculated patterns. (f) Temperature-dependent shift of the 2~$\theta$ values of (008) peak. }
\end{figure*}
 \begin{table*}[!t]
\caption{Comparison of lattice parameters, and unit-cell volume and $T_N$ for $\mathrm{CeCuBi_2^{os}}$ with reported values from Ref.~\cite{ye1996novel} and Ref.~\cite{jesus2014evolution}.} \label{T81}
\begin{centering}
\begin{tabular}{|l|l|l|l|l|}
\hline
\rowcolor{lightgray!30}\textbf{Sample} & \textbf{$a$=$b$(\textup {\AA})} & \textbf{$c$(\textup {\AA})} &  \textbf{Volume(\textup {\AA}$^3$)} &\textbf{$T_N$ (K)}\\ \hline
 \rowcolor{yellow!10}$\mathrm{CeCuBi_2^{os}}$   	         &4.552(7)&	9.650(8)	& 200.0(7)&14\\ \hline
\rowcolor{yellow!10}Ref~\cite{ye1996novel}: $\mathrm{CeCu_{0.7}Bi_2}$        &	4.555(2)  &	9.772(6)& 202.7(1)&-\\  \hline
 \rowcolor{yellow!10}Ref~\cite{jesus2014evolution}: $\mathrm{CeCuBi_2}$          &4.555(4)&	9.777(8) & 202.9(4)&16\\ \hline

\end{tabular}
\par\end{centering}
\end{table*}

In addition to its interesting magnetic properties, recent theoretical studies have proposed CeCuBi$_2$ as a possible candidate to realize multiple topological phases by the magnetic-field-induced spin reorientation~\cite{wang2021topologically}. These theoretical calculations suggest that the Cu square-net structure and hybridized electronic states near the Fermi level may provide an interesting setting to explore possible connections between correlated magnetism and electronic band topology in future studies~\cite{wang2021topologically}. However, experimental investigations of field-tuned electronic responses and angle-dependent transport properties in CeCuBi$_2$ remain largely unexplored.

In this work, we investigate off-stoichiometric $\mathrm{CeCuBi_2^{os}}$ (elemental composition determined nearly Ce:Cu:Bi = 1:0.8:2.2) single crystals grown by the Bi-flux method. Off-stoichiometry in correlated electron systems can subtly alter electronic hybridization, magnetic exchange interactions, and anisotropic responses, offering a possible route to tuning complex ground states. Motivated by the strongly anisotropic layered crystal structure and previous theoretical predictions for CeCuBi$_2$, we carried out comprehensive angle-dependent magnetization, magnetotransport, and AC susceptibility measurements. Our results reveal Kondo-driven weak heavy-fermion behavior, large room-temperature anisotropic and butterfly-like magnetoresistance, multiple field-induced metamagnetic phases, and glassy spin dynamics in intermediate field-induced states. These results uncover the complex interplay between electronic correlations, magnetic frustration, anisotropic transport and field-tunable magnetic states in CeCuBi$_2$, making it a possible platform for studying correlated anisotropic quantum phenomena.

\section{\label{sec:level2}EXPERIMENTAL DETAILS}

Single crystals of CeCuBi\textsubscript{2} were grown by Bi-flux method. High-purity Ce, Cu and Bi metal chunks were taken in atomic ratio 1:1:10 and loaded into an alumina crucible. The crucible was sealed inside an evacuated quartz ampoule under partial Ar atmosphere. The above materials were heated to 1100$^\circ$C and dwelled for 12 h, and then slowly cooled to 650$^\circ$C at a rate of 2$^\circ$C/h. At 650$^\circ$C, the excess flux was removed by centrifuging the ampoule at 650$^\circ$C.

The elemental composition of the crystals was determined using energy-dispersive X-ray spectroscopy (EDS) with a TESCAN Magna LMU system. Prior to EDS measurements, the crystal surfaces were polished to remove residual Bi flux, and point EDS measurements were performed at several locations on the crystal surface. Structural and phase characterization were carried out using powder X-ray diffraction (XRD) with Cu K$\alpha$ radiation on a Malvern PANalytical Empyrean diffractometer. For powder XRD measurements, the crystals were ground inside an Ar-filled glovebox and covered in a Mylar sheet to minimize oxidation. Back-reflection Laue diffraction measurements were performed using a PROTO LAUE-COS system. Heat-capacity measurements were carried out for $\mathrm{CeCuBi_2}$ and the diamagnetic reference compound LaCuBi\textsubscript{2}, which was synthesized under identical growth conditions.

Temperature-dependent DC magnetic susceptibility and magnetic-field-dependent magnetization measurements were performed using a 14 T Physical Property Measurement System (PPMS-CFMS, Cryogenic Ltd.). Angle-dependent magnetic measurements were carried out by manually rotating the crystal with respect to the applied magnetic field. Electrical resistivity and magnetoresistance measurements were performed using a 9~T PPMS (Quantum Design). Rotational anisotropic magnetoresistance (AMR) measurements were conducted using the rotator option of the PPMS. AC susceptibility measurements were performed using a 7 T MPMS3 system (Quantum Design).

\section{\label{sec:level3}RESULTS AND DISCUSSION}

\begin{figure*}
  \centering
  \includegraphics[width=0.8\textwidth]{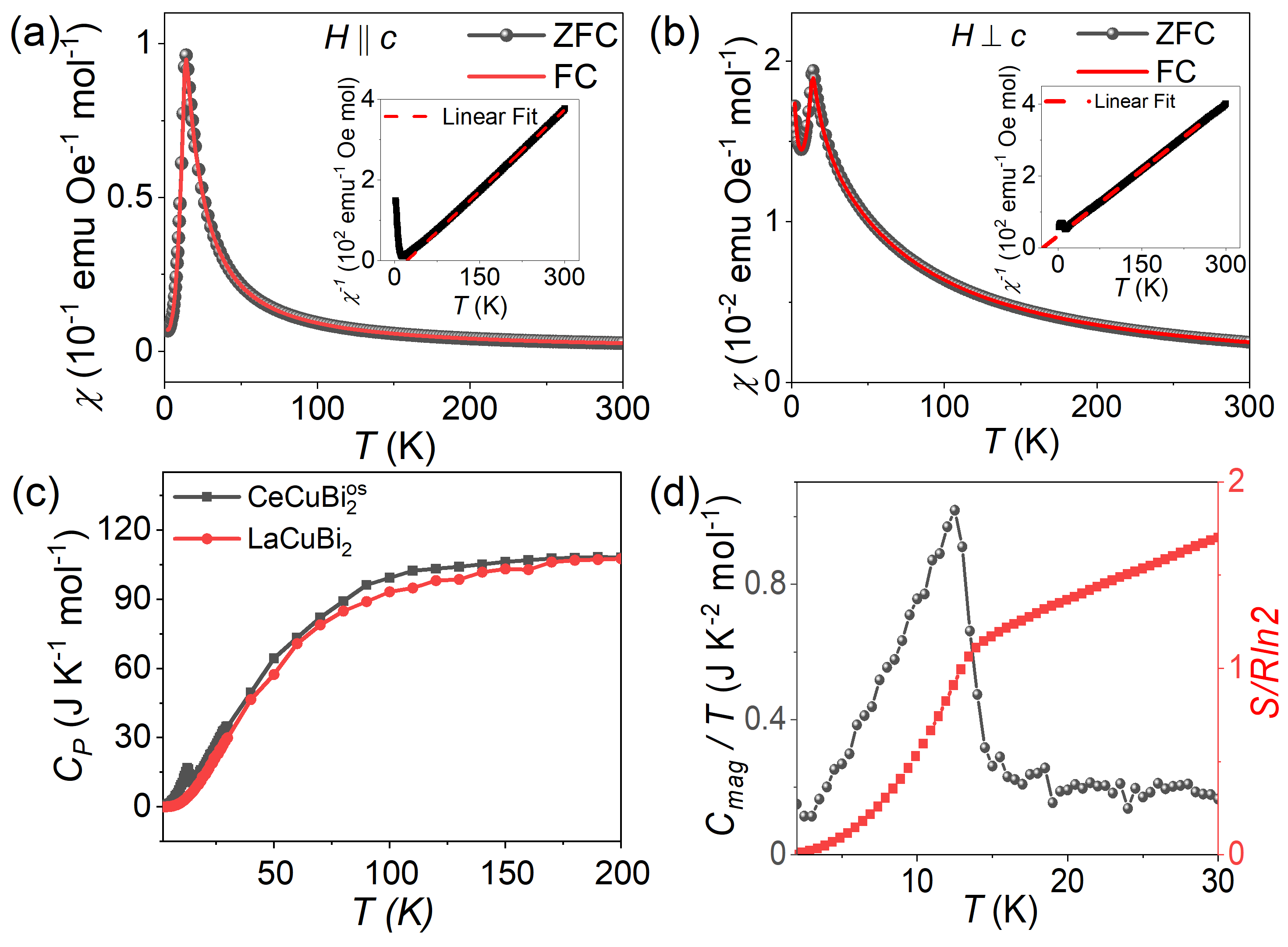}%
 
  \caption{\label{fig:four}
Temperature dependence of the zero-field-cooled (ZFC) and field-cooled (FC) DC magnetic susceptibility measured under an applied magnetic field of 100 Oe for (a) \(\textcolor{black}{H} \parallel [001]\) \((\chi_{\parallel})\) and (b) \(\textcolor{black}{H} \parallel [110]\) \((\chi_{\perp})\). Insets show the Curie--Weiss fitting of the inverse susceptibilities \(1/\chi_{\parallel}\) and \(1/\chi_{\perp}\) in the high-temperature paramagnetic region. (c) Temperature dependence of the specific heat \(C_P\) for $\mathrm{CeCuBi_2^{os}}$ and the diamagnetic reference compound LaCuBi$_{2}$. (d) Temperature dependence of the magnetic specific heat \(C_{\mathrm{mag}}/T\) and magnetic entropy \(S/R\ln2\) for \(\mathrm{CeCuBi_2^{os}}\).}
\end{figure*}
\begin{figure*}
    \centering
    \includegraphics[width=1\linewidth]{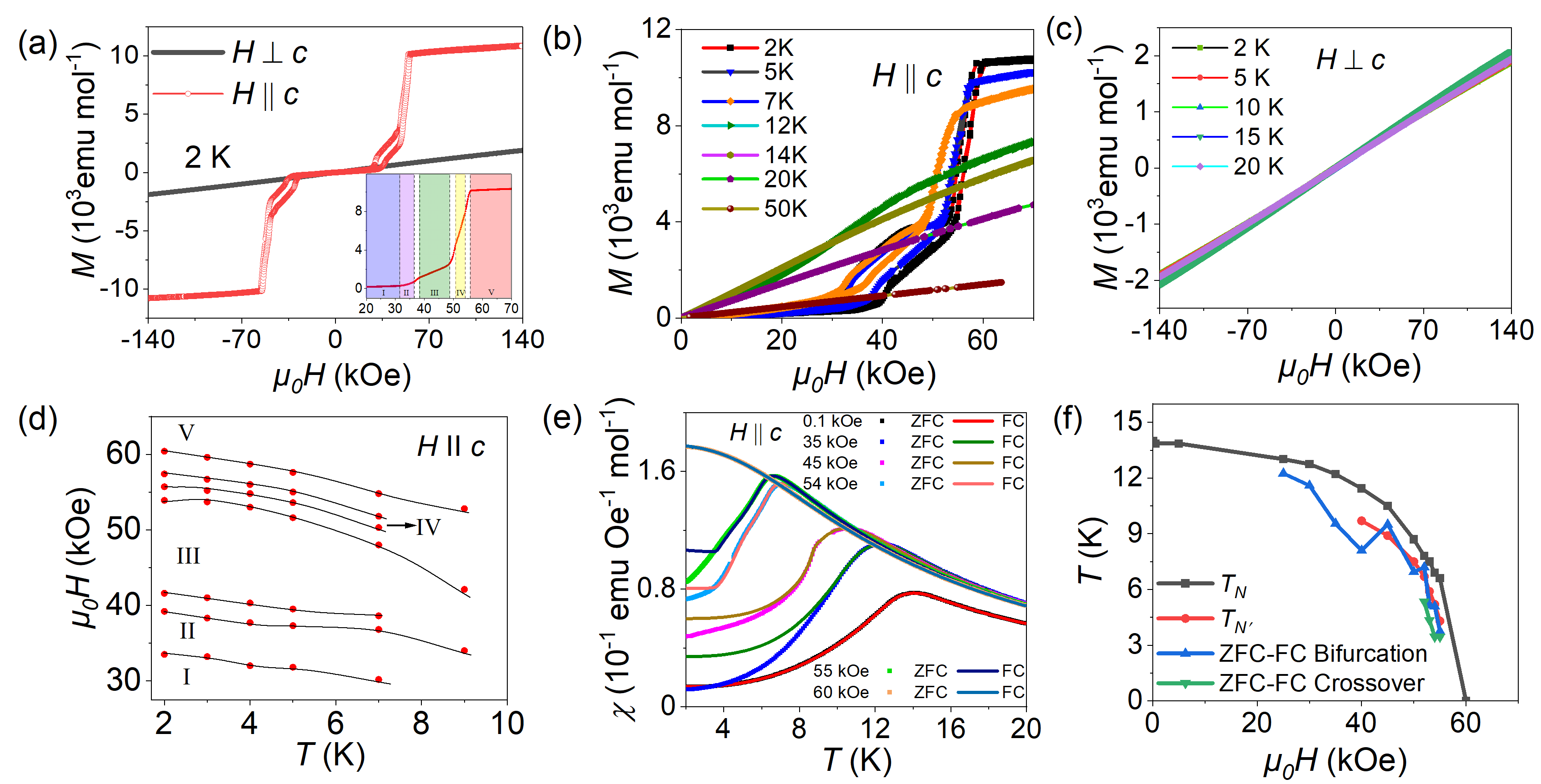}
    \caption{\label{fig:three}~(a) Field dependence of magnetization at 2~K for magnetic fields applied parallel and perpendicular to the $c$-axis.
   The inset highlights several field-induced magnetic phases and the multi-step evolution of magnetization. (b), (c) Magnetization curves measured at different temperatures for \(\textcolor{black}{H} \parallel c\) and \(\textcolor{black}{H} \perp c\), respectively. 
(d) Evolution of the metamagnetic transitions regions $I-V$ with temperature, calculated from field derivative of magnetization \textcolor{black}{$dM/dH$}
. (e) Temperature dependence of zero-field-cooled (ZFC) and field-cooled (FC) magnetic susceptibility measured for \(\textcolor{black}{H} \parallel c\) under different magnetic fields. (f) Magnetic field--temperature (\(\textcolor{black}{H}\)-\(T\)) phase diagram showing the evolution of \(T_N\), second transition \(T_{N'}\), ZFC--FC bifurcation, and crossover temperatures associated with the metamagnetic regime.}
   \end{figure*}

\textbf{Crystal Characterization:}~Figure~\ref{fig:one}~(a) shows the optical image of the as-grown CeCuBi$_2$ single crystal. The crystals are easily cleavable and exhibit a plate-like morphology, with dimensions of up to 4 mm x 5 mm x 1.5 mm. The crystal structure of CeCuBi$_2$ is shown in Fig.~\ref{fig:one}~(a), where Bi occupies two distinct sites (Bi1 and Bi2), while Ce atoms are responsible for the magnetic properties~\cite{adriano2014physical}. EDS measurements were performed on nine different regions of a freshly cleaved and polished crystal surface. The spectrum obtained from one representative region is displayed in Fig.~\ref{fig:one}~(b). The average elemental composition estimated from the EDS analysis is 1:0.8:2.2. The EDS results show that the crystals are off-stoichiometric with Cu deficiency and excess Bi. We have used the symbol $\mathrm{CeCuBi_2^{os}}$ throughout the manuscript to represent the off-stoichiometric (os) sample investigated in this work. $\mathrm{CeCuBi_2}$ has a layered structure, in which two-dimensional Bi layers consist of Bi atoms arranged in a square lattice array~\cite{ye1996novel}. Due to strong bonding between the Bi (atoms at Bi2 sites~\cite{ye1996novel}) and Cu atoms, the Bi2--Cu bond distance ($\sim$2.7~\AA) is significantly shorter than the sum of their metallic radii ($\sim$2.98~\AA). This gives rise to Cu vacancies in the lattice, commonly observed in the Ce$T$$X$$_2$ family of compounds. The Cu square net in $\mathrm{CeCuBi_2}$ occupies a position similar to that of the Si square net in ZrSiS, which may give rise to nodal lines in $\mathrm{CeCuBi_2}$~\cite{wang2021topologically}.

Figure~\ref{fig:one}~(c) shows the XRD pattern taken on the cleaved surface of the crystal. The peaks in the XRD pattern are indexed to (00l) planes, indicating that the crystal growth direction is along the $c$-axis. We have performed temperature-dependent XRD of the $\mathrm{CeCuBi_2^{os}}$ single crystal, the enlarged view of the (008) peak with temperature is shown in inset of Fig.~\ref{fig:one} (c). With increasing temperature, the (008) peaks are shifting towards lower 2$\theta$. This shift shows an increase in the lattice parameter $c$ due to thermal expansion. Figure~\ref{fig:one}~(d) shows the Laue diffraction pattern obtained from the crystal surface. The pattern is indexed with the (001) orientation, further confirming the single-crystalline nature of the sample.

Fig.~\ref{fig:one}~(e) shows the XRD pattern taken on crushed crystal powder. The $\mathrm{CeCuBi_2^{os}}$ powder was covered with a Mylar sheet inside a glovebox to avoid oxidation during measurements. Rietveld refinement of XRD data is performed using $P4/nmm$ space group. The lattice parameters of the tetragonal unit cell are compared with the reported values from Ref.~\cite{ye1996novel} and Ref.~\cite{jesus2014evolution}, as shown in~Tab.~\ref{T81}. The decrease in the $c$-lattice parameter and unit-cell volume is likely due to Cu vacancies. The vacancies at Cu sites and increase in pressure suppress the antiferromagnetic ordering temperature $T_N$ by modifying the $f$ ligand hybridization~\cite{piva2018high,seo2012pressure}. 
Further, no peak splitting is observed in the temperature-dependent XRD shown in the inset of Fig.~\ref{fig:one}~(c). The linear variation of the (008) peak position with temperature [Fig.~\ref{fig:one}~(f)] indicates the absence of a structural phase transition.

\textbf{DC Susceptibility and Specific Heat Measurements:}~Fig.~\ref{fig:four}~(a) and Fig.~\ref{fig:four}~(b) show the temperature-dependent zero-field-cooled (ZFC) and field-cooled (FC) DC magnetic susceptibility measured under an applied magnetic field of 100 Oe parallel to the $c$-axis [001] $\chi_{\parallel}$, and perpendicular to the $c$-axis along [110] $\chi_{\perp}$, respectively. Both $\chi_{\parallel}$ and $\chi_{\perp}$ exhibit antiferromagnetic ordering below the Néel temperature, $T_N \sim 14$ K. The overlap between the ZFC and FC curves indicates the absence of magnetic irreversibility~\cite{singh2026berry}. The ratio $\chi_{\parallel}$/$\chi_{\perp}$ $\approx$ 5 at $T_N$ is driven by tetragonal CEF splittings and provides a measure of Ce$^{3+}$ ions anisotropy. The large ratio indicates substantial CEF splitting between ground state and first excited state, along with large anisotropy of ground state and shows that $c$-axis is the easy axis of the magnetization~\cite{Thomas2016}. The Curie-Weiss law, $\frac{1}{\chi}=\frac{T-\theta_C}{C}$ was fitted to the high temperature linear paramagnetic region for 1/$\chi_{\parallel}$ and 1/$\chi_{\perp}$ as shown in the inset of Fig.~\ref{fig:four}~(a) and (b). The fitting yields Curie-Weiss temperatures of $\theta_{c_{\parallel}}$= 21.8 K, $\theta_{c_{\perp}}$=-27.5 K, \(\mu_{\mathrm{eff}}^{c\parallel} = 2.44\,\mu_{\mathrm{B}}/\mathrm{Ce}\) and \(\mu_{\mathrm{eff}}^{c\perp} = 2.56\,\mu_{\mathrm{B}}/\mathrm{Ce}\). The positive value of $\theta_{c_{\parallel}}$ indicates competing ferromagnetic interaction between the Ce ions along $c$-axis at higher temperatures, while the negative $\theta_{c_{\perp}}$ indicates dominant antiferromagnetic interactions within ab-plane.

The specific heat $C\textsubscript{P}$ as a function of temperature for $\mathrm{CeCuBi_2^{os}}$ and diamagnetic reference compound LaCuBi\textsubscript{2} is shown in Fig.~\ref{fig:four}~(c).  A sharp $\lambda$-type anomaly at $\sim$14 K indicates the antiferromagnetic ordering in this compound. The low-temperature $C/T$ data analyzed using the expression $C/T=\gamma + \beta T^{2}$, yielding $\gamma=102~\mathrm{mJ\,K^{-2}\,mol^{-1}}$ and 
$\beta=8~\mathrm{mJ\,K^{-4}\,mol^{-1}}$. In this relation, the linear 
$\gamma T$ term represents the electronic contribution to the specific heat, whereas the $\beta T^{3}$ term accounts for both lattice and magnon contributions. The enhanced value of Sommerfeld coefficient $\gamma=102~\mathrm{mJ\,K^{-2}\,mol^{-1}}$ in comparison to stoichiometric CeCuBi$_2$ (20) and CeAgBi$_2$ (52) indicates the weak heavy-fermion nature of the system~\cite{thamizhavel2003low}. Figure~\ref{fig:four}~(d) presents the magnetic specific heat $C_{\mathrm{mag}}/T$ of $\mathrm{CeCuBi_2^{os}}$ obtained by subtracting the lattice contributions from the diamagnetic reference compound. The temperature dependence of magnetic entropy was calculated by integrating $C_{\mathrm{mag}}/T$ with respect to temperature and is plotted as $S/R\ln2$ in Fig.~\ref{fig:four}~(d), where $R$ is the universal gas constant. At $T_N$ the entropy reached the value of $R\ln2$ which shows that CEF ground state is a doublet~\cite{thamizhavel2003low}. \textcolor{black}{The magnetic entropy continues to grow above $T_N$, indicating that the entropy associated with the Ce 4f moments is not released abruptly at the ordering temperature but is recovered over an extended temperature range. Possible causes include persistence of short range magnetic correlations above $T_{N}$, partial Kondo hybridization between Ce 4$f$ and conduction electrons that removes entropy gradually~\cite{Wang2019,Gornicka2026,Ashtar2026}, and thermal excitation of higher crystal electric field levels that contribute additional entropy at elevated temperatures~\cite{Wang2010,Chen2017}. In Kondo lattice systems the competition between Kondo screening and RKKY exchange often broadens the magnetic transition, so part of the entropy may be recovered gradually rather than at a single temperature~\cite{Ashtar2026}.}

\begin{figure*}[t]
  \centering
\includegraphics[width=\textwidth]{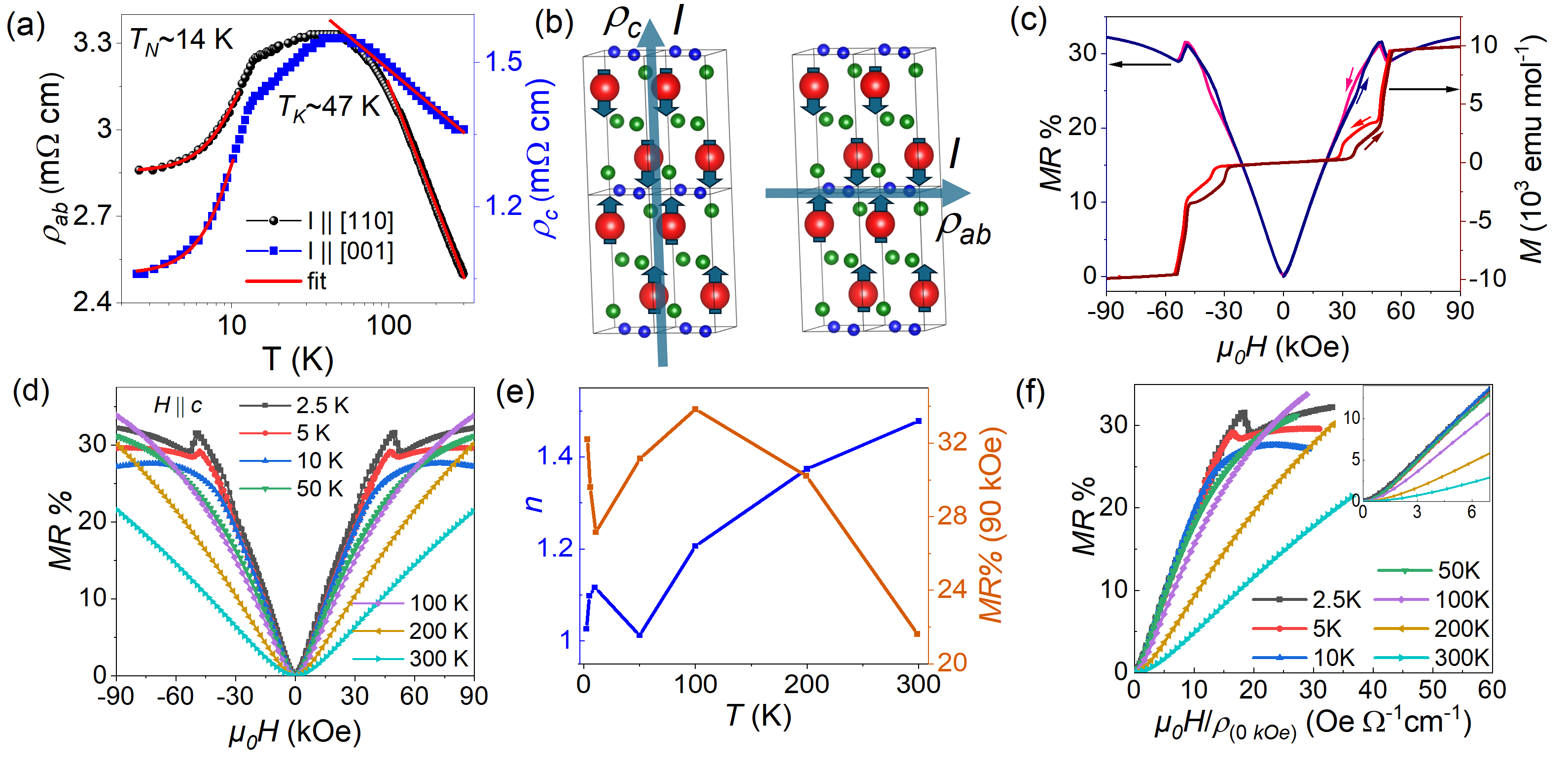}%
  
  \caption{\label{fig:eight}
 (a) Temperature dependence of longitudinal resistivity measured for current $I$ applied along the ab plane ([110]) $\rho_{ab}$ and along $c$-axis ([001]) $\rho_{c}$. The low-temperature resistivity data ($T$$\leq$10~K), fitted using Eq.~(1), resistivity has a maximum around $T$$\approx$47~K. Above, $T$$\approx$47~K, the resistivity follows a -$cln(T)$ dependence shown with linear red lines. (b) Schematic illustration of anisotropic scattering originating from crystallographic and magnetic anisotropy. The crystallographic anisotropy, determined by the orientation of Ce magnetic moments with respect to the current flow direction in the ab plane ([110]) and along the $c$-axis, can give rise to distinct high- and low-resistance states in $\rho_{ab}$ and $\rho_{c}$. The magnetic structure is adapted from Ref.~\cite{adriano2014physical}. (c) Magnetic field dependence of magnetoresistance MR\%={[$\rho_{c}$($\textcolor{black}{H}$)-$\rho_{c}$($0\textcolor{black}{\mathrm{kOe}}$)]/ $\rho_{c}$($0\textcolor{black}{\mathrm{kOe}}$)}$\times$100 measured at 2.5~K, together with the magnetization at 2~K. (d) Magnetic field dependence of MR\% at different temperatures. (e) Temperature dependence of exponent $n$, obtained from fitting $\mathrm{MR}\% = a\textcolor{black}{H}^{n}$ fit in low field region \textcolor{black}{(0--30~kOe)} 
 along with the MR\% at \textcolor{black}{90 kOe}
 .  (f) Kohler scaling of MR at different temperatures. The inset shows the low-field collapse of the MR curves and deviation from Kohler rule at higher temperature.}   
\end{figure*}

\textbf{Magnetization Measurements:}~Magnetization as a function of magnetic field at 2~K for fields applied parallel and perpendicular to the $c$-axis is shown in Fig.~\ref{fig:three}~(a). For $\textcolor{black}{H} \perp c$-axis, the magnetization increases linearly with field up to \textcolor{black}{140 kOe}, \textcolor{black}{without any metamagnetic transition.} The magnetization for applied magnetic field parallel to the $c$-axis remains linear up to $\textcolor{black}{\sim33 ~\mathrm{kOe}}$, and then exhibits multiple metamagnetic transitions and proceeds to increase in a multi-step process until complete spin-polarization near $\textcolor{black}{57 ~\mathrm{kOe}}$, marked by five different regions, as shown in inset of Fig.~\ref{fig:three}~(a). 

These metamagnetic transitions can be assigned to the spin-flop and spin-flip processes connected with field-induced rearrangements of the spin structure~\cite{feringa2022spin}. During these transitions, the spins gradually cant toward the direction of the applied magnetic field as the strength of field increases, ultimately reaching complete alignment at high fields. The hysteresis between the field up-sweep and down-sweep curves suggests first-order-like nature of these metamagnetic transitions.

The observed sequence of AFM, spin-flop, and fully polarized states result from the interplay among AFM exchange interactions, easy-axis anisotropy, uniaxial single-ion anisotropy, and Zeeman coupling for magnetic field applied along the AFM easy axis~\cite{li2016possible}. A spin-flop (SF) transition happens in an easy-axis antiferromagnet when the magnetic field along the easy-axis is stronger than the critical field $H_{\mathrm{SF}}=\sqrt{2H_AH_E-H_A^2}$, where $H_A$ is the anisotropy field and $H_E$ is the exchange field~\cite{feringa2022spin}. Above $H_{\mathrm{SF}}$, the spins reorient from a collinear antiferromagnetic state to a canted spin-flop configuration aligned toward the applied field\textcolor{black}{~\cite{Piva2020,Hodovanets2026,adriano2014physical,Thomas2016}}.

Magnetization curves with magnetic field applied parallel and perpendicular to the $c$-axis at different temperatures are shown in Fig.~\ref{fig:three}~(b) and Fig.~\ref{fig:three}~(c), respectively. For $\textcolor{black}{H} \parallel c$, the magnetization gradually evolves toward linear paramagnetic behavior with increasing temperature and becomes nearly paramagnetic above $T_N \approx 14$~K. Simultaneously, the metamagnetic states associated with the spin-flop transitions progressively weaken and disappear at higher temperatures. The magnetic fields corresponding to the spin-structure transitions for $\textcolor{black}{H} \parallel c$ are shown in Fig.~\ref{fig:three}~(d). 
At higher temperatures these metamagnetic transitions gradually weaken with increasing temperature, which may be attributed to the softening of the magnetic anisotropy~\cite{keffer1953spin,rathod2022electron} field $H_A$, leading to a reduction in the spin-flop field $H_{\mathrm{SF}}$.

Figure~\ref{fig:three} (e) shows the temperature dependent ZFC and FC magnetic susceptibility measured with the magnetic field applied parallel to the $c$-axis. With increasing magnetic field strength, $T_N$ gradually shifts toward lower temperatures, indicating suppression of the long-range antiferromagnetic order. In addition, a second anomaly $T_{N'}$ emerges in the field range \textcolor{black}{44--57~kOe} 
, suggesting the stabilization of an additional field-induced magnetic phase or a reconstruction of the spin configuration in the vicinity of the \textcolor{black}{metamagnetic transition~\cite{Li2026,lipika2026complex,singh2026berry}.} The field and temperature evolution of $T_N$ and $T_{N'}$ is summarized in the $\textcolor{black}{H}$-$T$ phase diagram shown in Fig.~\ref{fig:three}~(f). At low magnetic fields $(\textcolor{black}{\mu_{0}H < 22~\mathrm{kOe})}$, the ZFC and FC susceptibilities overlap, indicating reversible, equilibrium spin dynamics. However, at higher magnetic fields corresponding to the metamagnetic transition region, a clear bifurcation between the ZFC and FC curves develops. Such irreversibility is commonly associated with metastable spin configurations, enhanced domain-wall pinning, slow spin dynamics, or coexistence of competing magnetic phases induced by the applied field~\cite{kumar2021multiple}.  The corresponding bifurcation temperatures are mapped in the \(\textcolor{black}{H} \)-\(T\) phase diagram shown in Fig.~\ref{fig:three}~(f). 

Beyond the emergence of irreversibility, additional features appear at higher magnetic fields.
Interestingly, for applied fields of \textcolor{black}{45--56~kOe} 
, the FC curve initially falls below the ZFC curve and subsequently crosses it again at lower temperatures. This unusual ZFC--FC crossover and bifurcation may arise from nontrivial field-driven spin rearrangement~\cite{lipika2026complex}, phase coexistence, kinetically arrested metastable states, or reentrant magnetic behavior~\cite{chun2001reentrant}, where field cooling stabilizes a low-susceptibility canted spin configuration. These ZFC--FC crossover temperatures are also included in Fig.~\ref{fig:three}~(f). Furthermore at  \textcolor{black}{45 kOe and 56 kOe} 
, the FC curves become nearly flat and exhibit low-temperature saturation below \(T < 4\)~K, suggesting the formation of a metastable spin-locked state in the metamagnetic regime, where the spins become partially frozen or strongly polarized along the applied field direction. Here, thermal energy is insufficient to overcome anisotropy barriers, freezing the system into a field-polarized configuration with suppressed spin fluctuations.
 
\textbf{Resistivity and Magnetoresistance Measurements:}~The temperature dependence of longitudinal resistivity measured for current $I$ applied along the $ab$ plane ([110]) $\rho_{ab}$ and along $c$-axis $\rho_{c}$, is shown in Fig.~\ref{fig:eight}~(a). In both directions, $\rho_{ab}$ and $\rho_{c}$ exhibit a sharp change in slope at $T_N$ $\approx$ 14 K. Below $T_N$, the resistivity shows metallic behaviour with positive slope of temperature dependent resistivity, while above $T$ $\approx$ 47 K, resistivity decreases with a negative slope. \textcolor{black}{The broad resistivity maximum observed near 47 K is characteristic of Ce-based Kondo lattice compounds and is commonly associated with the crossover from incoherent Kondo scattering at high temperatures to coherent heavy-electron transport upon cooling~\cite{goltsev2005origin,Wang2019,feringa2022spin,Hodovanets2026,Li2026,singh2026berry,lipika2026complex,Samwer1976,Thompson1986,Lin1987}.} The temperature dependence of resistivity in the low temperature ($T$$<$$T_N$) antiferromagnetic metallic regime can be understood by using Matthiessen’s rule. In conventional antiferromagnets, the dominant scattering mechanisms contributing to the electrical resistivity are described by equation $\mathit{\rho}~(T)=\mathit{\rho}_{0}+\mathit{\rho}_{M}T^{3}+\mathit{\rho}_{P}T^{5}\text{-----(1)}$. Here, $\rho_{0}$ represents the temperature-independent residual resistivity arising from electron–defect scattering~\cite{rathod2022electron}. The coefficient $\rho_{M}$ is associated with the strength of electron–magnon scattering, which is expected to produce a  $T^3$ dependence in antiferromagnetic systems~\cite{ishikawa1982electrical}. Similarly, $\rho_{P}$ corresponds to the contribution from electron–phonon scattering, described by the Bloch–Grüneisen model, which gives rise to a $T^5$ dependence at temperatures well below the Debye temperature.  At low temperatures down to 2~K, the populations of magnons and phonons are strongly reduced, leading to a significant suppression of electron–magnon and electron–phonon scattering processes. As a result, the resistivity is predominantly governed by electron–defect scattering, which is generally expected to be nearly direction independent. The value of residual resistivity obtained in present work is significantly larger than those reported earlier for pristine CeCuBi$_2$, where $\rho_{ab}$ (2~K)= 0.024
m$\Omega$ cm~\cite{adriano2014physical}  and $\rho_{c}$ (2~K)= 0.073 m$\Omega$ cm~\cite{thamizhavel2003low}. The residual resistivity observed in the present compound is nearly two orders of magnitude higher than that reported for pristine CeCuBi$_2$~\cite{adriano2014physical, thamizhavel2003low}. This large enhancement is likely caused by strong electron-defect or disorder scattering originating from Cu vacancies or Bi interstitials in $\mathrm{CeCuBi_2^{os}}$.

Interestingly, $\mathrm{CeCuBi_2^{os}}$ exhibits huge anisotropy in the residual resistivity for current applied along different crystallographic directions with $\rho_{ab}$ (2~K)= 2.86 m$\Omega$ cm and $\rho_{c}$ (2~K)= 1.36 m$\Omega$ cm. The 
large difference between the zero-field residual resistivities measured along the two directions indicates that electron scattering is strongly influenced by the orientation of the antiferromagnetic moments relative to the current flow direction. As shown in the schematic of Fig.~\ref{fig:eight}~(b), the crystallographic anisotropy associated with the arrangement of Ce magnetic moments with respect to the current flow direction may give rise to distinct high and low-resistance states~\cite{singh2026berry}, leading to the observed difference between $\rho_{ab}$ (2~K) and  $\rho_{c}$ (2~K). The magnetic structure is adapted from Ref.~\cite{adriano2014physical}. In addition, the tetragonal crystal symmetry related anisotropic atomic arrangement within the layers may also contribute to the directional dependence of the resistivity. 

The low-temperature resistivity data ($T\leq$10~K) is fitted using eq. (1), and the corresponding fits are shown in Fig.~\ref{fig:eight}~(a). The magnitude of $\rho_{P}$ obtained from the fits are 1.18$\times$10$^{-9}$$~\Omega$~cm K$^{-5}$ and 1.24$\times$10$^{-9}$~$\Omega$~cm K$^{-5}$ for   $\rho_{ab}~(T)$ and $\rho_{c}~(T)$, respectively. Similarly, the fitted values of $\rho_{M}$ obtained for fit of $\rho_{ab}~(T)$ and $\rho_{c}~(T)$ are 3.43 $\times$ 10$^{-7}$~$\Omega$~cm K$^{-3}$ and 3.52 $\times$ 10$^{-7}$~$\Omega$~cm K$^{-3}$. The slightly larger value of $\rho_{M}$ along the $c$-axis suggests relatively stronger electron–magnon scattering in this direction, which is also consistent with the more pronounced drop in resistivity below $T_N$. At higher temperatures, the resistivity increases and forms a broad hump-like feature with a maximum around $T\approx$ 47~K and then drops monotonically at higher temperatures. To understand the origin of this broad feature, we consider Kondo scattering processes commonly observed in Ce-based correlated electron systems. Similar hump-like features have previously been reported in this family of compounds such as CeNiBi$_2$ and CeAgBi$_2$ which they attributed to the combined effect of the crystal electric field and the Kondo effect~\cite{thamizhavel2003low}.  In the paramagnetic state, localized Ce 4$f$ moments give rise to Kondo scattering, leading to a decrease in resistivity. Upon decreasing temperature, a broad hump-like feature appears, which may arise from the interplay among Kondo scattering, the development of weak heavy-fermion behavior, and the onset of magnetic ordering.~\cite{cornut1972influence,nicklas2001response}.

In pristine CeCuBi$_2$, a comparable high-temperature large decrease in resistivity was observed only under an applied pressure of about 18 kbar~\cite{adriano2014physical,piva2018high}. Previous studies have suggested that vacancies in the Cu layers of CeCuBi$_2$ can induce displacements of Cu atoms within the ab plane and Bi atoms along the $c$-axis~\cite{ye1996novel}. In $\mathrm{CeCuBi_2^{os}}$, the presence of Cu vacancies and excess Bi interstitials in the present compound could lead to a considerable microstrain and lattice disorder. Such structural changes can modify the Ce–Ce interatomic distance and therefore affect the Kondo coherence, resulting in pronounced hump-like features in the resistivity. 

The Kondo effect is usually observed as a minimum in the temperature dependent resistivity arising from scattering of conduction electrons by localized magnetic impurities.~\cite{kondo1964resistance}. For temperatures below this minimum, the resistivity follows a $-c\ln(T)$ dependence, where $c$ is related to the concentration of magnetic impurities~\cite{kondo1964resistance}. As the temperature approaches the Kondo temperature ($T_K$), this logarithmic behavior deviates and eventually gives rise to a maximum in resistivity, as observed in Fig.~\ref{fig:eight}~(a). In the present case, the presence of Cu vacancies and excess Bi interstitials could distort the magnetic environment of the neighboring Ce atoms, effectively creating localized magnetic impurity states within the $\mathrm{CeCuBi_2^{os}}$ lattice. Such defect-induced magnetic inhomogeneities may enhance the coupling between conduction electrons and localized moments, thereby strengthening Kondo interactions. In compounds with Ce impurities this hump-like feature originates from interplay between crystalline field and Kondo effect~\cite{cornut1972influence,hossain1999antiferromagnetic}.
\begin{figure*}[t]
  \centering
  \includegraphics[width=0.8\textwidth]{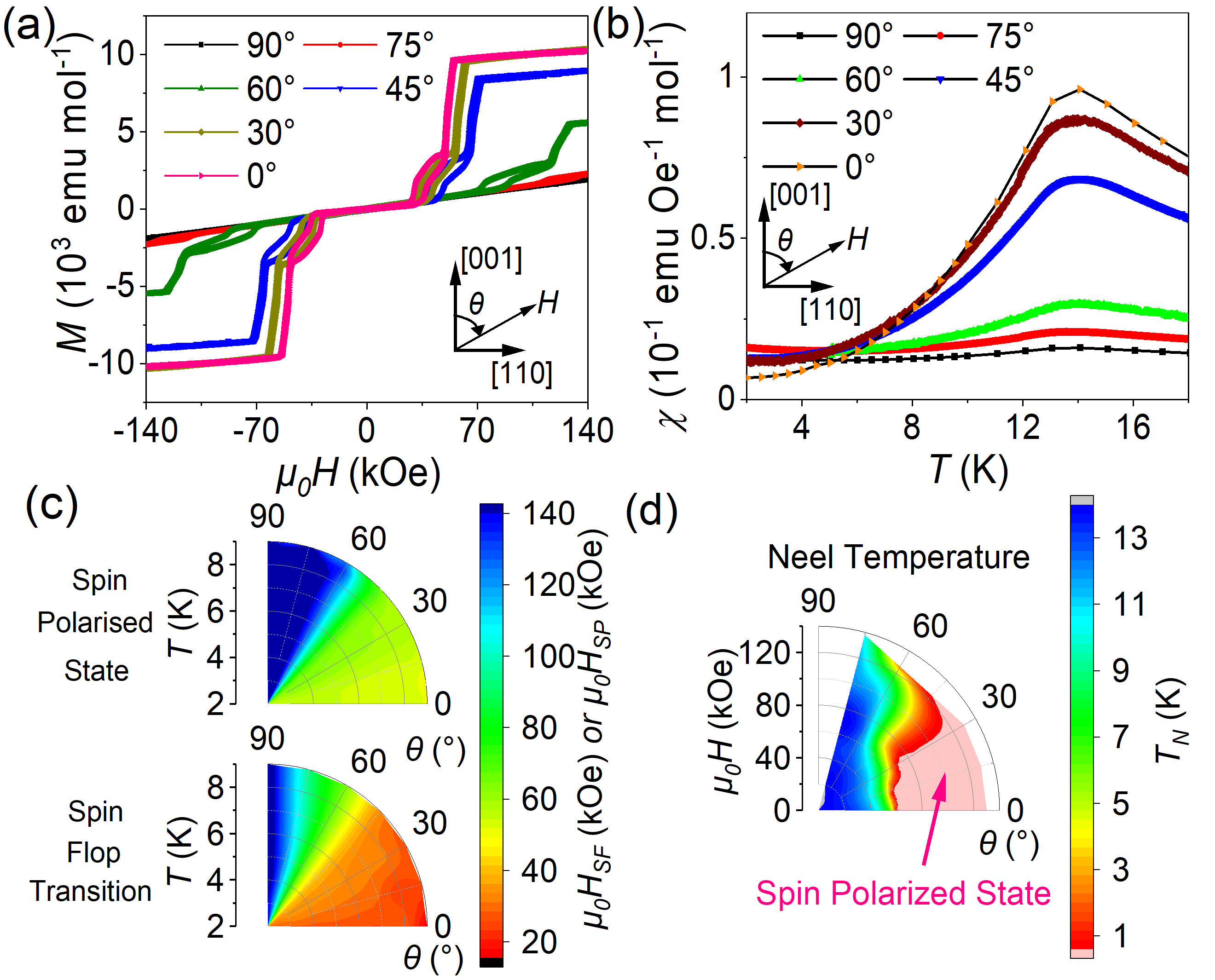}%

  \caption{\label{fig:ten}
  (a) Magnetic field dependence of the magnetization of $\mathrm{CeCuBi_2^{os}}$ measured at 2~K for different field orientations from $\textcolor{black}{H} \parallel [001]$ ($\theta = 0^\circ$) to $\textcolor{black}{H} \parallel [110]$ ($\theta = 90^\circ$). (b) Temperature dependence of the DC magnetic susceptibility for different field orientations from $\textcolor{black}{H} \parallel [001]$ ($\theta = 0^\circ$) to $\textcolor{black}{H} \parallel [110]$ ($\theta = 90^\circ$). (c) Polar phase diagrams showing the angular dependence of the fully spin-polarized field ($\textcolor{black}{H}_{SP}$) and the spin-flop transition field ($\textcolor{black}{H}_{SF}$) as a function of temperature. (d) Polar phase diagram illustrating the variation of the $T_N$ with applied magnetic field for different field orientations. 
  The color bar represents $T_N$, while the pink region corresponds to the fully spin-polarized state.}\end{figure*}

In most Kondo systems, the resistivity minimum and subsequent upturn occur at relatively low temperatures. However, in the present compound, Kondo scattering appears to be unusually strong, resulting in a broad semiconducting-like hump extending to much higher temperatures. Such behavior is often observed in various Ce-based Kondo insulators due to a hybridization gap arising from Kondo coherence~\cite{xiang2021unusual,dzero2012theory,chen2015magnetoresistance,RAUCHSCHWALBE1987347,hundley1993magnetoresistance}. Nevertheless, the comparatively low resistivity at high temperature $\rho_{ab}$ (300~K)= 2.5 m$\Omega$ cm relative to $\rho_{ab}$ (2~K)= 2.86 m$\Omega$ cm suggests that the observed behavior is dominated by enhanced Kondo scattering rather than carrier hopping mechanisms. Furthermore, as shown in Fig.~\ref{fig:eight}~(a), the larger logarithmic slope coefficient $c$ for $\rho_{ab}~(T)$ compared to $\rho_{c}~(T)$ indicates stronger magnetic-impurity-related scattering along the $c$-axis. Kondo coupling enhances due to doping or pressure related lattice contraction and enhancement of RKKY interactions~\cite{mendoncca2008tuning,adriano2014physical}. \textcolor{black}{The comparatively stronger Kondo scattering observed for current flowing along the c-axis may reflect anisotropic hybridization between the localized Ce 4f electrons and conduction electrons associated with the layered crystal structure\cite{muro2013anisotropic}. Additionally, the comparatively stronger $-clnT$ slope observed in $\mathrm{CeCuBi_2^{os}}$ than in pristine CeCuBi$_2$ is possibly associated with enhanced Kondo hybridization resulting from the reduction in the lattice parameter $c$ caused by Cu vacancies\cite{goltsev2005origin}.} 

Figure~\ref{fig:eight} (c) shows the field dependence of magnetoresistance (MR\%) and magnetization of $\mathrm{CeCuBi_2^{os}}$ for a magnetic field $\textcolor{black}{H}$ applied along the $c$-axis at 2~K. At \textcolor{black}{90 kOe} 
, we observe a large MR$\sim$32\%. Both MR and magnetization exhibit hysteresis upon increasing and decreasing the magnetic field. At low magnetic fields, the MR is positive and increases linearly with field. A decrease in MR emerges near 
$\textcolor{black}{H}$$~\sim$~\textcolor{black}{49 - 53 kOe} 
, coinciding with a sharp rise in magnetization. This behavior reflects a strong suppression of spin-disorder scattering as the spins align under the applied field. With further increase in the magnetic field, the MR gradually approaches saturation, closely tracking the behavior of the magnetization,  which indicates that the MR is strongly coupled to the field-evolved magnetic configuration.

Figure~\ref{fig:eight} (d) show the field dependence of MR\% at different temperatures. As the temperature increases from 2.5~K, the MR initially decreases. Interestingly, at 100~K the MR reaches its largest value $\sim$ 34\% at \textcolor{black}{90 kOe} 
. Despite the substantial positive magnetoresistance observed in the present compound, its magnitude remains lower than that reported for closely related rare-earth silver antimonide and rare-earth dibismuthide compounds~\cite{myers1999systematic,petrovic2002anisotropic,petrovic2003anisotropic}. Furthermore, a large room-temperature magnetoresistance of about 22\% is observed at 300 K and \textcolor{black}{90 kOe}. 
The low-field MR \textcolor{black}{(0-30 kOe)}
~at different temperatures is fitted using $\mathrm{MR}\% = a\textcolor{black}{H}^n$. The values of the exponent $n$ obtained from the fitting, along with the MR\% at \textcolor{black}{90 kOe} 
, are shown in Fig.~\ref{fig:eight}~(e). \textcolor{black}{Such a phenomenological power-law analysis is widely used to parameterize the magnetic-field dependence of magnetoresistance and quantify deviations from the conventional quadratic ($H^2$) orbital behavior. The field dependence of MR has been analyzed in RPtBi, LuPtBi, LuPdBi, VAs$_2$, and Cr$_2$NiGa, revealing linear or nearly quadratic behavior associated with carrier compensation, high mobility, and semiclassical transport~\cite{Ruvalds1988,Sampathkumaran,Nakatsuji2004,Pikul2012,Hodovanets2015,Hu2008,Pavlosiuk2015}.} The value of $n$ is close to 1 at low temperatures ($T \leq 10$ K) and deviates at higher temperatures. Linear MR can arise either from linear electronic bands~\cite{abrikosov1998quantum} or from mobility fluctuations associated with inhomogeneities~\cite{parish2003non} possibly associated with Cu vacancies. 

To gain further insight into the underlying transport mechanisms, we analyzed the magnetoresistance using Kohler scaling~\cite{dasoundhi2024extremely}. The Kohler scaling of MR at different temperatures is shown in Fig.~\ref{fig:eight}~(f). At low fields \textcolor{black}{($\mu_0H<~$30 kOe)} 
, the low temperature (2.5-50~K) MR curves collapse onto a single curve, i.e., they follow Kohler's rule. This suggests that, at low temperatures (2.5-50~K) and low fields \textcolor{black}{($\mu_0H<~$30 kOe)} 
, the charge carriers at the Fermi surface possess similar scattering times. However, Kohler's rule is violated at high fields as well as at higher temperatures above 50~K, implying the presence of multiple scattering mechanisms or field-induced modifications of the electronic structure. The enhanced MR around 100~K may be associated with small field-induced modifications of the Fermi surface, a possible field induced Lifshitz transition, or the breakdown of Kondo coherence~\cite{onishi2018deviation}.

\begin{figure*}[t]
  \centering
  
  \includegraphics[width=\textwidth]{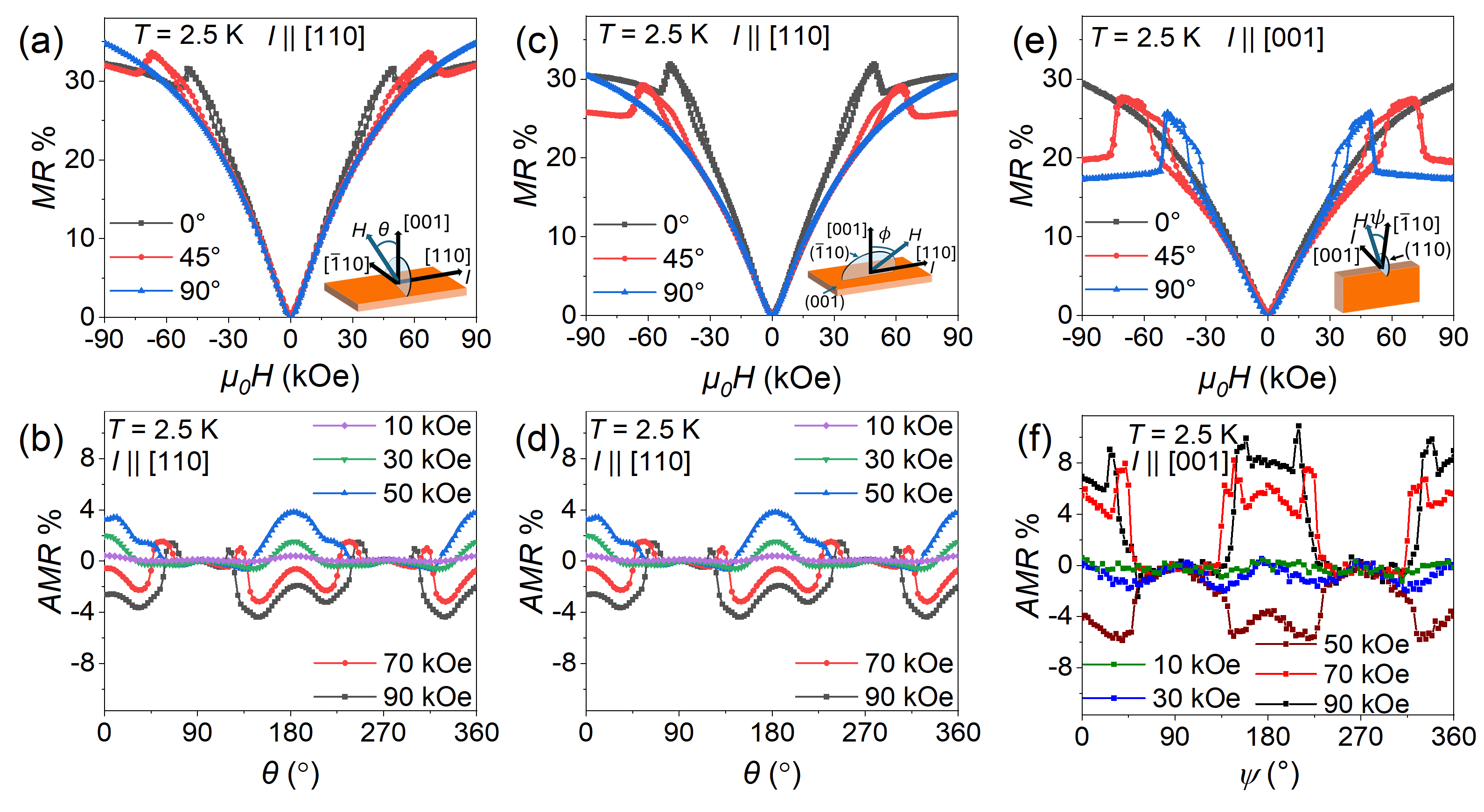}%
 
    \caption{\label{fig:eleven}
    (a) Magnetic field dependence of MR\% at 2.5~K with current along the $ab$-plane ([110] direction); the magnetic field is rotated from $\textcolor{black}{H} \parallel [001]$ ($\theta = 0^\circ$) to $\textcolor{black}{H} \parallel [\bar{1}10]$ ($\theta = 90^\circ$), as shown in the schematic. (b) Angular dependence of AMR\% measured in the same configuration as in panel (a), with $\theta$ varied from $0^\circ$ to $360^\circ$: \( \mathrm{AMR}\,\% = \frac{\rho(\theta)-\rho(\theta=90^\circ)} {\rho(\theta=90^\circ)} \times 100 \). (c) Magnetic field dependence of MR\% at 2.5~K with current along the $ab$-plane ([110] direction); the magnetic field is rotated from $\textcolor{black}{H} \parallel [001]$ ($\phi = 0^\circ$) to $\textcolor{black}{H} \parallel [110]$ ($\phi = 90^\circ$), as shown in the schematic. (d) Angular dependence of AMR\% measured in the same configuration as in panel (c), with $\phi$ varied from $0^\circ$ to $360^\circ$.(e) Magnetic field dependence of MR\% at 2.5~K with current along the $c$-axis; the magnetic field is rotated from  $\textcolor{black}{H} \parallel [\bar{1}10]$ ($\psi = 0^\circ$) to $\textcolor{black}{H} \parallel [001]$ ($\psi = 90^\circ$), as shown in the schematic. (f) Angular dependence of AMR\% measured in the same configuration as in panel (e), with $\psi$ varied from $0^\circ$ to $360^\circ$.}
\end{figure*}

To sum up, the large MR observed over the entire temperature range (2--300 K) arises from interplay of two dominant scattering mechanisms. In the antiferromagnetic state ($T < T_N$), where the resistivity follows magnon-dominated behavior, the appearance of decreasing MR near the metamagnetic transition \textcolor{black}{($\mu_0H \sim 49$--$53$ kOe at 2~K)} 
indicates a key role of spin-disorder scattering. In contrast, at higher temperatures ($T > T_K$), where semiconducting/Kondo-insulator-like resistivty behavior~\cite{xiang2021unusual,dzero2012theory,chen2015magnetoresistance,RAUCHSCHWALBE1987347} is observed suggest that positive MR is possibly driven by the magnetic-field-induced breakdown of Kondo coherence~\cite{hossain1999antiferromagnetic,hundley1993magnetoresistance}. Thus, the overall MR in $\mathrm{CeCuBi_2^{os}}$ reflects the combined effects of positive contributions from Kondo-coherence breakdown and negative contributions from reduced spin-disorder scattering. \textcolor{black}{This interpretation is further supported by the tendency of both the magnetization and the magnetoresistance to saturate at sufficiently high magnetic fields and low temperatures, in agreement with the progressive stabilization of the field-polarized magnetic state. At higher temperatures the thermal fluctuations suppress the field-induced spin polarization, such that the full saturation is not reached in the investigated field range.}

\textbf{Angle Dependent Magnetization Measurements:}~Figure~\ref{fig:ten}~(a) shows the magnetic field dependence of the magnetization of $\mathrm{CeCuBi_2^{os}}$ measured at 2~K for different field orientations, varying from $\textcolor{black}{H}
\parallel [001]$ ($\theta = 0^\circ$) to $\textcolor{black}{H}
\parallel [110]$ ($\theta = 90^\circ$). The magnetization behavior for $\textcolor{black}{H}
\parallel [001]$ ($\theta = 0^\circ$) and $\textcolor{black}{H}
\parallel [110]$ ($\theta = 90^\circ$) has already been discussed in Fig.~\ref{fig:three}~(a). The metamagnetic spin-flop transition occurring at \textcolor{black}{$\sim$29~kOe for $H \parallel c$} 
shifts to higher fields and gradually weakens with increasing angle, eventually disappearing for $\textcolor{black}{H} \perp c$ 
. This suggests that, at low angles, the metamagnetic transitions are associated with the large magnetocrystalline anisotropy along the $c$-axis, whereas at higher angles away from anisotropy axis the magnetic moments undergo a gradual canting process instead of abrupt spin alignment. Figure~\ref{fig:ten}~(b) shows the temperature dependence of the DC magnetic susceptibility for different field orientations, varying from $\textcolor{black}{H} \parallel [001]$ 
($\theta = 0^\circ$) to $\textcolor{black}{H} \parallel [110]$ 
($\theta = 90^\circ$). The susceptibility for $\textcolor{black}{H} \parallel [001]$ 
($\theta = 0^\circ$) and $\textcolor{black}{H} \parallel [110]$ 
($\theta = 90^\circ$) is discussed in Fig.~\ref{fig:three}~(b) and (c). The antiferromagnetic ordering temperature, $T_N \sim 14$ K, remains unchanged for all field orientations due to the strong exchange interactions between neighboring Ce spins that stabilize the long-range antiferromagnetic order. In contrast, the magnitude of susceptibility at $T_N \sim 14$ K decreases as the field is rotated away from the easy anisotropy axis. The polar phase diagrams constructed from magnetization measurements performed at different temperatures and magnetic field orientations are shown in Fig.~\ref{fig:ten}~(c). These diagrams illustrate the angular dependence of the magnetic fields required to achieve the fully spin-polarized state ($\textcolor{black}{H}_{SP}$) 
and the spin-flop (S-F) transition field ($\textcolor{black}{H}_{SF}$) 
as a function of temperature. Figure~\ref{fig:ten}~(d) presents the polar phase diagram, illustrating how the Néel temperature varies with applied field at different angles.

\begin{figure*}[t]
  \centering
  \includegraphics[width=\textwidth]{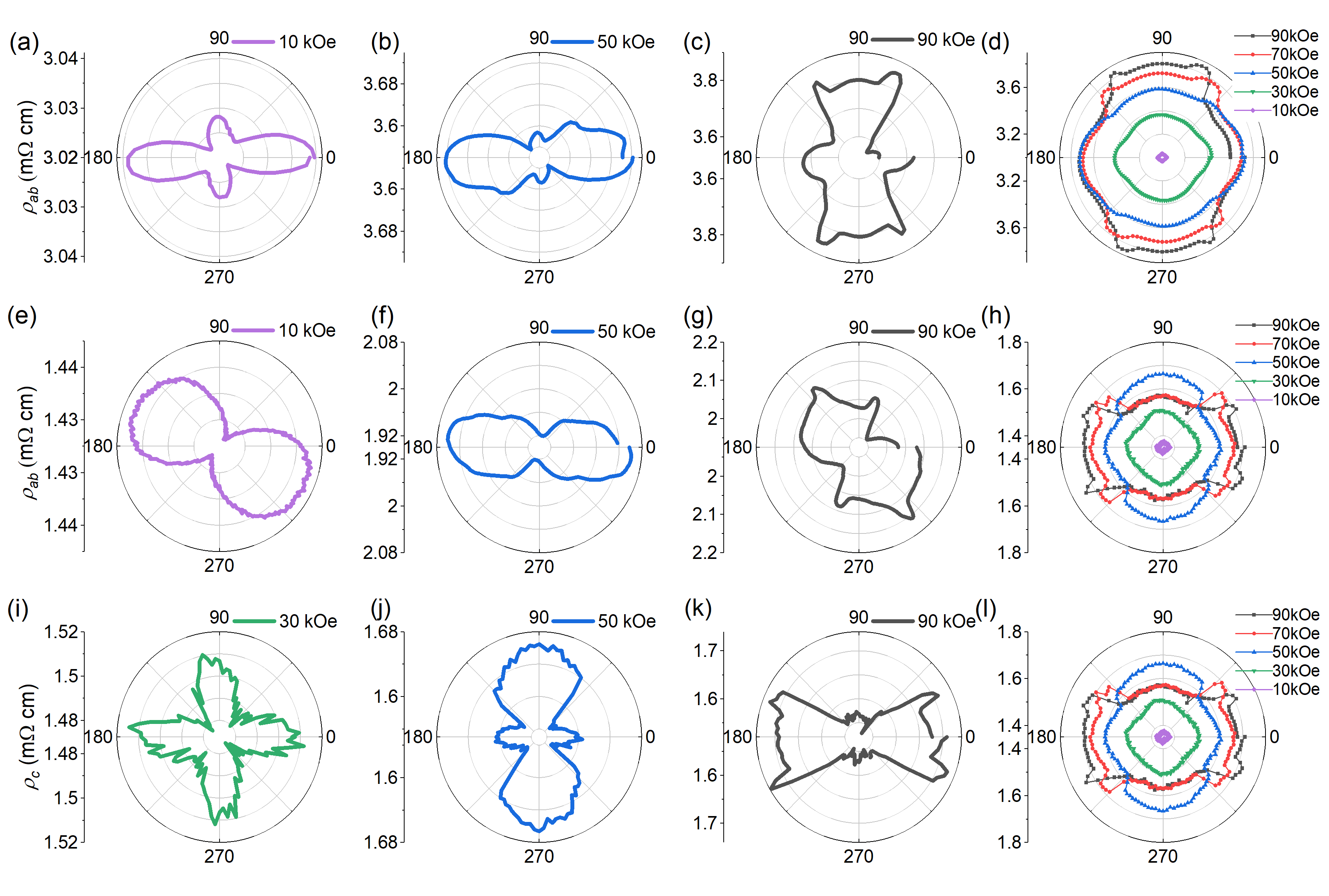}
  
  \caption{
  \label{fig:twelve}
  Polar plots of angle dependent resistivity measured at 2.5~K with the magnetic field rotated from $0^\circ$ to $360^\circ$. For current applied along the $ab$-plane ([110] direction), the magnetic field is rotated from $\textcolor{black}{H} \parallel [001]$ ($\theta = 0^\circ$) to $\textcolor{black}{H} \parallel [\bar{1}10]$ ($\theta = 90^\circ$) at constant magnetic fields of \textcolor{black}{(a) 10~kOe, (b) 50~kOe, (c) 90~kOe}, and (d) combined data from \textcolor{black}{10--90~kOe}. Similarly the magnetic field is rotated from $\textcolor{black}{H} \parallel [001]$ ($\phi = 0^\circ$) to $\textcolor{black}{H} \parallel [110]$ ($\phi = 90^\circ$) at constant magnetic field of \textcolor{black}{(e) 10~kOe, (f) 50~kOe, (g) 90~kOe}, and (h) combined data for \textcolor{black}{10--90~kOe}. For current applied along the $c$-axis, the magnetic field is rotated from $\textcolor{black}{H} \parallel [\bar{1}10]$ ($\psi = 0^\circ$) to $\textcolor{black}{H} \parallel [001]$ ($\psi = 90^\circ$) at magnetic fields of \textcolor{black}{(i) 30~kOe, (j) 50~kOe, (k) 90~kOe}, and (l) combined data from \textcolor{black}{10--90~kOe}.
  }
\end{figure*}

\textbf{Angle Dependent Magnetoresistance Measurements:}~Magnetic field dependence of the MR\% measured at 2.5~K with the current applied along the $ab$-plane ([110] direction), is shown in Fig.~\ref{fig:eleven}~(a). The magnetic field is rotated from $\textcolor{black}{H} \parallel [001]$ 
($\theta = 0^\circ$) to $\textcolor{black}{H} \parallel [\bar{1}10]$ 
($\theta = 90^\circ$).  The MR\% for $\textcolor{black}{H} \parallel [001]$ 
was discussed earlier in Fig.~\ref{fig:eight}~(c). As shown in Fig.~\ref{fig:eleven}~(a), for $\theta = 0^\circ$, the MR reaches $\sim$32\% at \textcolor{black}{90 kOe} 
. The MR is initially positive and nearly linear with field, while a negative MR appears near $\textcolor{black}{H} \sim \textcolor{black}{49}$--$\textcolor{black}{53}$~\textcolor{black}{kOe} 
, coinciding with a sharp increase in magnetization. At higher fields, the MR gradually saturates, following the behavior of magnetization. For $\theta = 45^\circ$ the MR\% at \textcolor{black}{90~kOe} 
remains nearly unchanged; however, the field corresponding to the sharp negative MR shifts to higher values. For $\textcolor{black}{H} \parallel [\bar{1}10]$ 
($\theta = 90^\circ$), the MR\% increases to $\sim$35\%. The enhanced MR is likely associated with weaker suppression of spin disorder along [110], where the magnetization response remains nearly linear and unsaturated. Fig.~\ref{fig:eleven}~(b) shows the AMR\% measured for the same configuration as in Fig.~\ref{fig:eleven}~(a), exhibiting a maximum positive AMR of approximately 3.8\% near $\theta = 0^\circ$ and $180^\circ$ at \textcolor{black}{50~kOe} 
.
 A similar trend is observed when the magnetic field is rotated from $\textcolor{black}{H} \parallel [001]$ 
 ($\phi = 0^\circ$) to $\textcolor{black}{H} \parallel [110]$ 
 ($\phi = 90^\circ$), as shown in Fig.~\ref{fig:eleven}~(c). Fig.~\ref{fig:eleven}~(d) shows the AMR\% measured for the same configuration as in Fig.~\ref{fig:eleven}~(c), exhibiting a maximum positive AMR of approximately 7.2\% near $\phi = 0^\circ$ and $180^\circ$ at \textcolor{black}{50~kOe} 
 . Figure~\ref{fig:eleven}~(e) shows the MR for current applied along $c$-axis [001] with the magnetic field rotated from $\textcolor{black}{H} \parallel [\bar{1}10]$ 
 ($\psi = 0^\circ$) to $\textcolor{black}{H}\parallel [001]$ 
 ($\psi = 90^\circ$). At $\psi = 0^\circ$ the MR\% $\sim$ 30\% at \textcolor{black}{90~kOe} 
 and decreases with increasing field angle. Interestingly, for both $\psi = 45^\circ$ and $\psi = 90^\circ$, the MR exhibits a sharp enhancement of about 8\%, followed by a subsequent decrease of nearly 8\%. This indicates a field-driven reconfiguration of the magnetic spin structure associated with the metamagnetic transition. The initial enhancement corresponds to a high resistance magnetic state that raises resistivity, while the subsequent drop signals partial spin alignment (reduced scattering) as the moments settle into a more polarized configuration.  Fig.~\ref{fig:eleven}~(f) presents the AMR\% measured for the same configuration as in Fig.~\ref{fig:eleven}~(e), exhibiting a maximum positive AMR of approximately 10.9\% near $\psi = 0^\circ$ and $180^\circ$ at \textcolor{black}{90~kOe} 
 . Among the three AMR configurations, the largest positive AMR\%$\sim$ 10.9\% is observed along with negative AMR\%$\sim$ 6\% at \textcolor{black}{50 kOe} 
 , when the current is applied along the easy $c$-axis and the magnetic field is rotated in the (110) plane, from $\textcolor{black}{H} \parallel [\bar{1}10]$ 
 ($\psi = 0^\circ$) to $\textcolor{black}{H} \parallel [001]$ 
 ($\psi = 90^\circ$).

This configuration also shows sudden variations of resistivity, which indicates a strongly anisotropic magnetoresistance response. The resistance can also show sudden jumps, due to the magnetic domain wall pinning/unpinning during the magnetization reversal or a metamagnetic transition~\cite{ritzinger2023anisotropic,bolte2005magnetotransport}. Reported values of AMR in antiferromagnets are FeRh~(1\%), CuMnAs~(0.3\%), CeAlGe$_{0.72}$Si$_{0.28}$~(5\%) and Sr$_2$IrO$_4$~($\sim$160\%)~\cite{marti2014room,wang2019giant,yang2021colossal}. The positive AMR observed in our antiferromagnetic $\mathrm{CeCuBi_2^{os}}$ sample reaching $\sim$10.9~\% is significant and puts it among antiferromagnetic systems with large AMR responses~\cite{ritzinger2023anisotropic}. Recently EuMnSb$_2$ and EuTe$_2$ displayed colossal AMR near a field-driven transition~\cite{soh2019magnetic,yang2021colossal}.

The AMR effect is generally explained using the \(s\)-\(d\) scattering model together with spin--orbit coupling, where the resistivity depends on the angle between the current and the magnetization/N\'eel vector, typically following a \(\cos2\theta\) dependence. In antiferromagnets, AMR originates from the field-induced reorientation of the N\'eel vector, which modifies the electronic band structure and carrier scattering processes~\cite{ritzinger2023anisotropic}. A small negative AMR component was also observed, which may arise from dominant same-spin scattering processes (\(s\uparrow \rightarrow d\uparrow\) or \(s\downarrow \rightarrow d\downarrow\)), changes in the spin-polarized density of states near the Fermi level, or magnetic disorder effects~\cite{ritzinger2023anisotropic}. The large AMR observed in $\mathrm{CeCuBi_2^{os}}$ indicates strong coupling between spin orientation, spin--orbit interaction, and charge transport. The enhanced magnitude suggests significant anisotropic scattering, slight deformation of the Fermi surface due to rotation of magnetic moments, and possible crystalline contributions to the AMR. Furthermore, band structure calculations for spin-polarized CeCuBi$_2$ suggest the presence of spin-minority bands, mainly contributed by Cu \(s\)-orbitals, near the Fermi level with tilted band crossings along the \(\Gamma\)--X and M--\(\Gamma\) directions~\cite{wang2021topologically}. Such electronic structure features may influence the anisotropic transport properties and could contribute to the unusual AMR behavior observed at high fields.

\begin{figure*}
    \centering
    \includegraphics[width=1\linewidth]{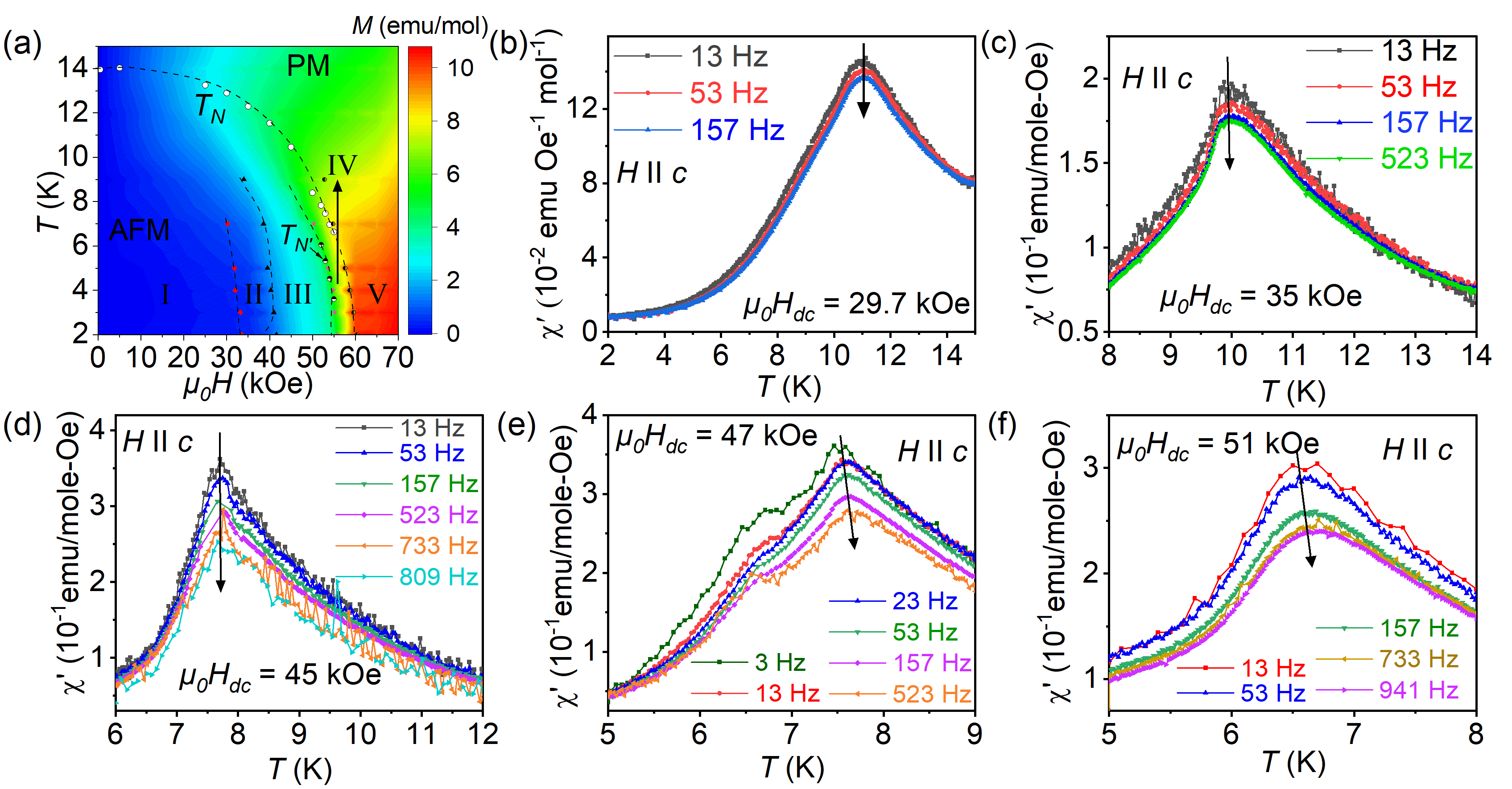}
    \caption{(a) Magnetic field–temperature (\textit{\textcolor{black}{H}--T}) phase diagram highlighting the different field-induced magnetic phases (I--V) and the paramagnetic (PM) region. Temperature dependence of the real part of the \textcolor{black}{AC} susceptibility $\chi'(\textit{T})$, measured at different excitation frequencies using an \textcolor{black}{AC} field of amplitude 4.5 Oe with a superimposed \textcolor{black}{DC} magnetic field of (b) \textcolor{black}{29.7 kOe}, (c) \textcolor{black}{35 kOe}, (d) \textcolor{black}{45 kOe}, (e) \textcolor{black}{47 kOe}, and (f) \textcolor{black}{51 kOe}.}
    \label{fig:placeholder}
\end{figure*}

\begin{figure*}[t]
    \centering
    \includegraphics[width=0.8\linewidth]{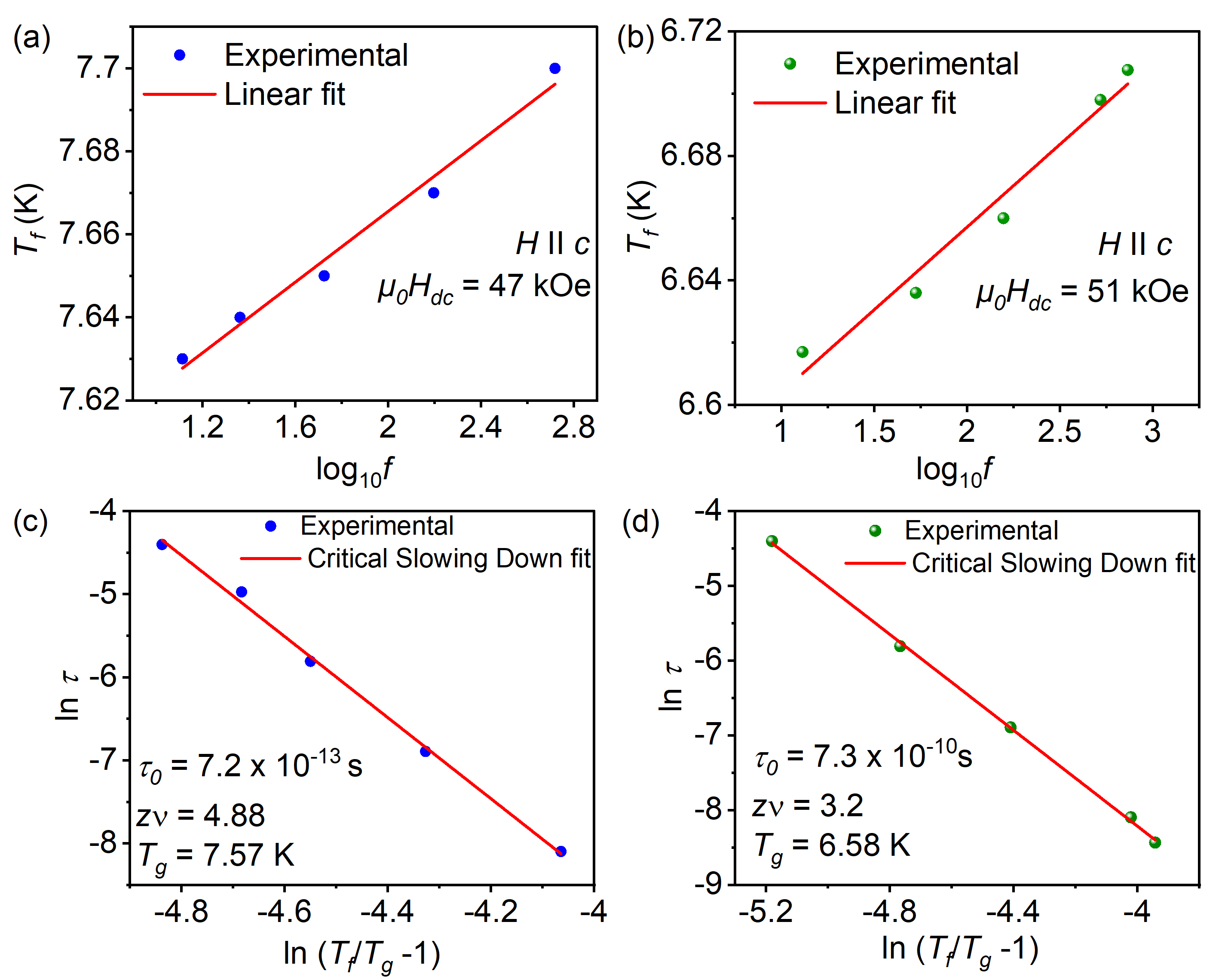}
    \caption{(a)The plot of spin-freezing temperature $T_f$ vs $\log_{10}f$ obtained at \textcolor{black}{DC} field of \textcolor{black}{47 kOe} and (b) \textcolor{black}{51 kOe}. Plot of $\ln\tau$ vs $\ln(T_f/T_g-1)$, fitted using the  critical slowing down model for (c) \textcolor{black}{47 kOe} and (d) \textcolor{black}{51 kOe}.}
    \label{fig:89}
\end{figure*}

\textbf{Polar Plots of Angle-Dependent Resistivity Measurements:}~Figure~\ref{fig:twelve} displays polar plots of the angle-dependent resistivity measured at 2.5~K as the magnetic field is rotated through a full angular range from $0^\circ$ to $360^\circ$. Panels Fig.~\ref{fig:twelve}~(a)–(d) correspond to measurements with the current applied in the $ab$ plane along [110], while the magnetic field is rotated from $\textcolor{black}{H}\parallel[001]$ 
($\theta=0^\circ$) to $\textcolor{black}{H}\parallel[\bar{1}10]$ 
($\theta=90^\circ$) at fixed magnetic fields of 10, 50, and 90~kOe 
, respectively, along with the combined \textcolor{black}{10--90~kOe}
~dataset. At \textcolor{black}{10 kOe}
~[Fig.~\ref{fig:twelve}~(a)], the polar plot exhibits four-lobed structure with an overall twofold symmetry. The observed twofold symmetry implies periodic repetition after a $180^\circ$ rotation, consistent with the equivalence of the magnetic field orientations along the [001] and [00$\bar{1}$] directions and the low field enlargement suggesting dominant anisotropic scattering associated with the underlying magnetic configuration along $c$-axis. As the magnetic field increases to \textcolor{black}{50 kOe} 
~[Fig.~\ref{fig:twelve}~(b)], the pattern expands primarily along the horizontal direction, accompanied by a symmetric displacement of the enhanced lobes. The enhancement may be due to a gradual field-induced modification of the magnetic structure. At \textcolor{black}{90 kOe} 
[Fig.~\ref{fig:twelve}~(c)], the angular profile transforms into an eight-lobed butterfly-like structure. This indicates the highly anisotropic nature of the magnetic and electronic structure under a \textcolor{black}{90 kOe} 
field. The sharp changes observed at \textcolor{black}{90 kOe} 
upon varying the field angle may also arise from metamagnetic transitions, higher-order crystalline contributions, and the presence of small two-dimensional Fermi surface features. Figure~\ref{fig:twelve}~(d) summarizes the field evolution using the combined \textcolor{black}{10--90~kOe}
~dataset.

Panels Fig.~\ref{fig:twelve}~(e)–(h) present measurements for the same in-plane current configuration, but with the magnetic field rotated from $\textcolor{black}{H}\parallel[001]$ 
($\varphi=0^\circ$) to $\textcolor{black}{H}\parallel[110]$
($\varphi=90^\circ$). At \textcolor{black}{10 kOe}
~[Fig.~\ref{fig:twelve}~(e)], the polar plot consists of two rounded lobes, forming a dumbbell-like pattern tilted by approximately $45^\circ$ relative to the horizontal axis, suggesting a preferred conduction direction. At \textcolor{black}{50 kOe}
~[Fig.~\ref{fig:twelve}~(f)], the pattern expands and becomes a more pronounced dumbbell-like structure, reflecting enhanced field-driven anisotropy. Upon increasing the magnetic field to \textcolor{black}{90 kOe} 
[Fig.~\ref{fig:twelve}~(g)], two additional lobes emerge, while maintaining the underlying twofold symmetry. The appearance of additional lobes indicates the involvement of higher-order anisotropic components and field-dependent electronic structure evolution. Figure~\ref{fig:twelve}~(h) displays the corresponding combined \textcolor{black}{10--90~kOe} 
dataset.  

Panels Fig.~\ref{fig:twelve}~(i)--(l) show measurements with current applied along the $c$-axis while the magnetic field is rotated from $\textcolor{black}{H}\parallel[\bar{1}10]$ 
($\psi=0^\circ$) to $\textcolor{black}{H}\parallel[001]$ 
($\psi=90^\circ$). At \textcolor{black}{30 kOe}
~[Fig.~\ref{fig:twelve}~(i)], the polar plot exhibits a four-lobed structure with four-fold symmetry. This suggests that, at \textcolor{black}{30 kOe}  
field, the magnetic structure and/or Fermi surface exhibit symmetry along the [001], [00$\bar{1}$] and [$\bar{1}$10] and [110] directions. At \textcolor{black}{50 kOe} 
[Fig.~\ref{fig:twelve}~(j)], the pattern expands predominantly along the vertical direction, possibly due to field-induced redistribution of magnetic structure and related scattering. At \textcolor{black}{90 kOe} 
[Fig.~\ref{fig:twelve}~(k)]
, the angular dependence again evolves into an eight-lobed butterfly-like structure. The data points at $0^\circ$ and $360^\circ$ do not completely overlap, suggesting possible hysteretic behavior associated with magnetization or domain reorientation effects. ZrSiS crystallizes in the same space group as CeCuBi$_2$ and hosts nodal-line features in its electronic structure that resemble the spin-polarized state proposed for CeCuBi$_2$ under a magnetic field of \textcolor{black}{90 kOe} 
, where possible nodal-line features may emerge~\cite{wang2021topologically}. 

The butterfly-shaped anisotropic magnetoresistance in ZrSiS has been understood by the collective effects of the Zeeman-tuned nearly perfect electron-hole compensation and the highly anisotropic Fermi surface of tubular hole pockets and large electron pockets. The interplay between these factors gives rise to the butterfly-like angular magnetoresistance~\cite{voerman2019origin}. Overall, the angular dependence of the resistivity is strongly anisotropic, indicating pronounced sensitivity to the magnetic field orientation, and suggests a close interplay between magnetic order, Fermi surface topology and anisotropic charge transport.

\textbf{AC susceptibility measurements:}~A magnetic field–temperature (\textit{\textcolor{black}{H}–T})
~phase diagram is constructed from temperature and field-dependent DC magnetization measurements, as shown in Fig.~\ref{fig:placeholder} (a). The corresponding field-induced magnetic phases I–V are represented with a color gradient indicating the variation in magnetization strength across the different phases. At constant fields of \textcolor{black}{29.7 kOe and 30 kOe} 
, cooling below the $T_N$ drives phase transitions from the PM state to phases I and II, respectively. Similarly, at \textcolor{black}{45 kOe}
, a transition from the PM state to phase III occurs upon decreasing the temperature. To investigate the relaxation dynamics associated with the different magnetic phases, the $\chi'(\textit{T})$ data were measured at various excitation frequencies under a superimposed \textcolor{black}{DC} magnetic field and an \textcolor{black}{AC} field with an amplitude of 4.5 Oe, as shown in Fig.~\ref{fig:placeholder}~(b). At constant DC fields (\textit{$\textcolor{black}{H}_{dc}$})
of \textcolor{black}{29.7 kOe, 35 kOe, and 45 kOe} 
, corresponding to regions I, II, and III, respectively, no noticeable frequency dependence of the transition peak is observed, as displayed in Fig.~\ref{fig:placeholder} (b), (c), and (d). This behavior suggests the absence of slow relaxation processes within the investigated frequency range, indicating that long-range magnetic order remains preserved across these magnetic phases.

In contrast, for $\textcolor{black}{H}_{dc}$ = \textcolor{black}{47 kOe} and $\textcolor{black}{H}_{dc}$ = \textcolor{black}{51 kOe}
, as illustrated in Fig.~\ref{fig:placeholder}~(e) and Fig.~\ref{fig:placeholder}~(f), a pronounced shift of the peak toward higher temperatures is observed with increasing excitation frequency. The corresponding field values and associated peak temperatures fall within region IV of the \textit{\textcolor{black}{H}–T}
~phase diagram. These peak positions in $\chi'(\textit{T})$ correspond to the spin-freezing temperature, $T_f$. At $\textcolor{black}{H}_{dc}$ = \textcolor{black}{47 kOe} and $\textcolor{black}{H}_{dc}$ = \textcolor{black}{51 kOe}
, as shown in Fig.~\ref{fig:89} (a) and Fig.~\ref{fig:89} (b), $T_f$ exhibits a linear shift with $\log_{10}~(f)$. Such frequency-dependent behaviour is indicative of slow spin dynamics and suggests the emergence of a glassy magnetic state in this region.

To quantify the observed peak shift, we calculated the Mydosh parameter, defined as $K$=$\Delta T_f/(T_f\Delta log_{10}f)$, where $\Delta T_f$ is the shift in freezing temperature $T_f$, and \textit{f} is the excitation frequency~\cite{mydosh1993spin}. The values of $K$ for \textcolor{black}{47 kOe} and \textcolor{black}{51 kOe} 
are 0.006 and 0.008 respectively. Both values lie within the range 0.005 to 0.01 reported for canonical spin-glass systems~\cite{mulder1981susceptibility,giot2008magnetic}. To estimate the spin relaxation time ($\tau_0$), we have applied critical slowing down model, which is given as,  $\tau = \tau_0 (T_f/T_g-1)^{-z\nu}$, where $\tau$=1/2$\pi f$ is the relaxation time, $\tau_0$ is the relaxation time corresponding to an individual spin flip, $z\nu$ is the dynamical critical exponent, $T_f$ is the spin freezing temperature for frequency \textit{f} and $T_g$ is the freezing temperature for zero frequency limit~\cite{mydosh1993spin,manna2011magnetic,guo2016spin,PhysRevB.91.224420,manna2014correspondence}. The best fit of this model for \textcolor{black}{47 kOe}
~[Fig.~\ref{fig:89}~(c)] and \textcolor{black}{51 kOe}
~[Fig.~\ref{fig:89}~(d)] corresponds to a relaxation time ($\tau_0$) of 7.2 x $10^{-13}$ s and 7.3 x $10^{-10}$ s, respectively. These values of $\tau_0$ lie in the range of [$10^{-10}$- $10^{-13}$] reported for typical canonical spin glass systems~\cite{chakrabarty2014cluster, maji2011low}. The spin-glass-like parameters obtained above further motivate discussion of the microscopic origin of the field-induced glassiness. The observed slow spin dynamics in the field induced magnetic state in region IV at \textcolor{black}{47 kOe}
~and \textcolor{black}{51 kOe}
~suggest that the existence of mixed ferro-antiferromagnetic interactions are inducing magnetic frustration, resulting in spin glass-like slow spin relaxation in  $\mathrm{CeCuBi_2^{os}}$ single crystals~\cite{feng2006glassy,hiroi2009ferromagnetism}.\\

\section{\label{sec:discussion}DISCUSSION}
\textcolor{black}{The physical properties of off-stoichiometric CeCuBi$_2$ can be understood in a unified framework governed by the competition between Kondo screening, RKKY exchange interaction, crystalline electric field anisotropy and magnetic disorder induced by Cu vacancies~\cite{seo2012pressure}. The enhanced Sommerfeld coefficient and the wide resistivity maximum indicate a moderate Kondo hybridization coexisting with long-range antiferromagnetic order below $T_N$, which places $\mathrm{CeCuBi_2^{os}}$ in the weak heavy-fermion regime~\cite{thamizhavel2003low}. At the same time, the Cu vacancies may modify the local electronic environment around the Ce ions and modify the hybridization between the localized 4f electrons and the conduction electrons, leading to a distribution of exchange interactions and magnetic anisotropies~\cite{goltsev2005origin,slebarski2020role}. The competition between the exchange interaction, anisotropy and Zeeman energy stabilizes successive field induced magnetic phases under the magnetic field applied along the easy $c$-axis~\cite{ishii2021field}. The resulting field-induced spin configurations strongly influence the scattering of the conduction electrons leading to the pronounced anisotropic magnetoresistance, butterfly-like AMR and the close correspondence between the magnetic and magnetotransport anomalies. We therefore demonstrate that the magnetic and transport responses are different manifestations of the same correlated electronic state, which is dominated by the competition between Kondo hybridization, magnetic anisotropy, and field-tunable spin configurations. Although the present measurements provide a good correlation between the magnetic and transport properties, the microscopic magnetic structures of the intermediate phases need to be determined by field-dependent neutron diffraction or resonant X-ray magnetic scattering.}




\section{\label{sec:conclusion}CONCLUSION}

In conclusion, we report structural, magnetic, angle‑dependent magnetotransport, and spin‑dynamics measurements on off‑stoichiometric CeCuBi$_2$ single crystals. The compound orders antiferromagnetically below $T_N\sim14$ K with strong anisotropy and weak heavy-fermion behavior with Sommerfeld coefficient $\gamma =102$ mJ K$^{-2}$ mol$^{-1}$. Transport studies show highly anisotropic resistivity and a broad hump near $\sim47$ K, indicating Kondo-effect-driven weak heavy-fermion behavior possibly associated to Cu vacancies and Ce magnetic disorder. The pronounced anisotropy observed in magnetization, electrical transport, and magnetoresistance likely originates from the quasi-two-dimensional layered magnetic structure, which strongly influences the electronic and magnetic interactions. A large room-temperature magnetoresistance of $\sim$22\% at 300~K and \textcolor{black}{90 kOe}
, together with butterfly-like anisotropic magnetoresistance reaching $\sim$10.9\% at 2.5~K and \textcolor{black}{90 kOe}
, demonstrates strong coupling among magnetic order, anisotropic scattering, and spin--orbit interactions. We observe multiple field-induced metamagnetic phases, AC susceptibility measurements at \textcolor{black}{51 kOe}
~reveal field-induced glassy spin dynamics with a Mydosh parameter of $K$ = 0.008 and relaxation time of $\tau_0$~$\sim$7.3$\times$$10^{-10}$~s, indicating the emergence of a field-induced canonical spin-glass-like magnetic state. These findings establish off-stoichiometric CeCuBi$_2$ as a versatile platform where Kondo-driven heavy-fermion behavior, large anisotropic magnetotransport, and field-induced glassiness coexist, making it a possible candidate for exploring correlated and anisotropic quantum phenomena.

\section*{ACKNOWLEDGMENTS}

K.M. acknowledges financial support from the Max Planck Society under the Max Planck--India Partner Group Project; the Science and Engineering Research Board (SERB), Department of Science and Technology (DST), Government of India, through Grant No.~CRG/2022/001826; the Aeronautics Research and Development Board (ARDB, Project No.~1992); and the Defence Research and Development Organisation (DRDO) under Project No.~DFTM/033203/P/41/JATC-P2QP17/10/D(R\&D)/2023. V.C. acknowledges the Ministry of Human Resource Development (MHRD) for Institute Fellowship support and funding through the GIMRT Program of the Institute for Materials Research, Tohoku University (Proposal No.~ 202308-CRKKE-0509). S.R. acknowledges the Anusandhan National Research Foundation (ANRF), India, for support through the National Postdoctoral Fellowship under Grant No.~PDF/2025/003558.
The authors also acknowledge the Central Research Facility (CRF) and the Nanoscale Research Facility at IIT Delhi for providing access to materials characterization facilities, including PPMS, MPMS, EDX, and XRD measurements.

\bibliography{apssamp}

@article{ishikawa1982electrical,
  title={Electrical Resistivity Due to Antiferromagnetic Spin Waves in {Cr}},
  author={Ishikawa, Akio},
  journal={J. Phys. Soc. Jpn.},
  volume={51},
  number={2},
  pages={441--451},
  year={1982},
  doi = {10.1143/JPSJ.51.441},
  publisher={The Physical Society of Japan}
}

@article{goltsev2005origin,
  title={Origin of the pressure dependence of the Kondo temperature in Ce-and Yb-based heavy-fermion compounds},
  author={Goltsev, AV and Abd-Elmeguid, MM},
  journal={J. Phys.: Condens. Matter},
  doi = {10.1088/0953-8984/17/11/011},
  volume={17},
  number={11},
  pages={S813--S821},
  year={2005}
}

@article{muro2013anisotropic,
  title={Anisotropic c--f hybridization in the Kondo semiconductor {CeFe$_2$Al$_10$}},
  author={Muro, Yuji and Yutani, Keiske and Kajino, Jumpei and Onimaru, Takahiro and Takabatake, Toshiro},
  journal={J. Korean Phys. Soc.},
  doi = {10.3938/jkps.63.508},
  volume={63},
  number={3},
  pages={508--511},
  year={2013},
  publisher={Springer}
}

@article{ishii2021field,
  title={Field-induced successive phase transitions in the J 1-J 2 buckled honeycomb antiferromagnet {Cs$_3$Fe$_2$Cl$_9$}},
  author={Ishii, Y and Narumi, Y and Matsushita, Y and Oda, M and Kida, T and Hagiwara, M and Yoshida, HK},
  journal={Phys. Rev. B},
  doi = {10.1103/PhysRevB.103.104433},
  volume={103},
  number={10},
  pages={104433},
  year={2021},
  publisher={APS}
}

@article{slebarski2020role,
  title={Role of Ce 4 f-conduction band on-site hybridisation in the nature of magnetism in {Ce$_5$MGe$_2$}, where M is d-electron-type metal},
  author={{\'S}lebarski, Andrzej},
  journal={Philos. Mag.},
  doi = {10.1080/14786435.2019.1671620},
  volume={100},
  number={10},
  pages={1193--1203},
  year={2020},
  publisher={Taylor \& Francis}
}

@article{rathod2022electron,
  title={Electron-magnon scattering in an anisotropic half-metallic ferromagnetic Weyl semimetal {Co$_3$Sn$_2$S$_2$}},
  author={Rathod, Shivam and Malasi, Megha and Lakhani, Archana and Kumar, Devendra},
  journal={Phys. Rev. Mater.},
  volume={6},
  number={8},
  pages={084202},
  year={2022},
  doi = {10.1021/acs.chemmater.4c01384},
  publisher={APS}
}

@article{singh2026berry,
  title={Berry curvature induced giant anomalous and spin texture driven Hall responses in the layered kagome antiferromagnet {GdTi$_3$Bi$_4$}},
  author={Singh, Shobha and Rathod, Shivam and Chen, Rong and Lipika and Sneh and Umetsu, Rie Y and Sun, Yan and Manna, Kaustuv},
  journal={Phys. Rev. B},
  volume={113},
  number={13},
  pages={134437},
  year={2026},
  doi={10.1103/h8sf-m1zv},
  publisher={APS}
}

@article{thamizhavel2003low,
  title={Low temperature magnetic properties of {CeTBi$_2$} {(T: Ni, Cu and Ag)} single crystals},
  author={Thamizhavel, Arumugam and Galatanu, Andrei and Yamamoto, Etsuji and Okubo, Tomoyuki and Yamada, Mineko and Tabata, Kanehito and C Kobayashi, Tatsuo and Nakamura, Noriko and Sugiyama, Kiyohiro and Kindo, Koichi and others},
  journal={J. Phys. Soc. Jpn.},
  volume={72},
  number={10},
  pages={2632--2639},
  year={2003},
  doi={10.1143/JPSJ.72.2632},
  publisher={The Physical Society of Japan}
}

@article{adriano2014physical,
  title={Physical properties and magnetic structure of the intermetallic {CeCuBi$_2$}  compound},
  author={Adriano, Cris and Rosa, Priscila Ferrari Silveira and Jesus, Camilo BR and Mardegan, Jos{\'e} Renato Linares and Garitezi, Thales Macedo and Grant, Taran and Fisk, Z and Garcia, Daniel Julio and Reyes, AP and Kuhns, PL and others},
  journal={Phys. Rev. B},
  volume={90},
  number={23},
  pages={235120},
  year={2014},
  doi={10.1103/PhysRevB.90.235120},
  publisher={APS}
}

@article{piva2018high,
  title={High-pressure studies on heavy-fermion antiferromagnet {CeTBi$_2$}},
  author={Piva, M{\'a}rio Moda and Ajeesh, MO and Christovam, DS and Dos Reis, RD and Jesus, CBR and Rosa, Priscila Ferrari Silveira and Adriano, C and Urbano, RR and Nicklas, M and Pagliuso, PG},
  journal={J. Phys.: Condens. Matter},
  volume={30},
  number={37},
  pages={375601},
  year={2018},
  doi={https://doi.org/10.1088/1361-648x/aad7d8},
  publisher={IOP Publishing}
}

@article{kondo1964resistance,
  title={Resistance minimum in dilute magnetic alloys},
  author={Kondo, Jun},
  journal={Prog. Theor. Phys.},
  volume={32},
  number={1},
  pages={37--49},
  year={1964},
  doi={10.1143/PTP.32.37},
  publisher={Oxford University Press}
}

@article{nicklas2001response,
  title={Response of the heavy-fermion superconductor {CeCoIn$_5$} to pressure: roles of dimensionality and proximity to a quantum-critical point},
  author={Nicklas, M and Borth, R and Lengyel, E and Pagliuso, PG and Sarrao, JL and Sidorov, VA and Sparn, G and Steglich, F and Thompson, JD},
  journal={J. Phys.: Condens. Matter},
  volume={13},
  number={44},
  pages={L905--L912},
  year={2001},
  doi={10.1088/0953-8984/13/44/104}
}

@article{cornut1972influence,
  title={Influence of the crystalline field on the {Kondo} effect of alloys and compounds with cerium impurities},
  author={Cornut, B and Coqblin, B},
  journal={Phys. Rev. B},
  volume={5},
  number={11},
  pages={4541},
  year={1972},
  doi={10.1103/PhysRevB.5.4541},
  publisher={APS}
}

@article{hossain1999antiferromagnetic,
  title={Antiferromagnetic {Kondo}-lattice systems {Ce$_2$Rh$_3$Ge$_5$} and {Ce$_2$Ir$_3$Ge$_5$} with moderate heavy-fermion behavior},
  author={Hossain, Z and Ohmoto, Hirohisa and Umeo, Kazunori and Iga, Fumitoshi and Suzuki, Takashi and Takabatake, Toshiro and Takamoto, Naoki and Kindo, Koichi},
  journal={Phys. Rev. B},
  volume={60},
  number={14},
  pages={10383},
  year={1999},
  doi={10.1103/PhysRevB.60.10383},
  publisher={APS}
}

@article{ye1996novel,
  title={A Novel Intermetallic Compound, {CeCu$_{1-x}$Bi$_2$}},
  author={Ye, J and Huang, YK and Kadowaki, K and Matsumoto, T},
  journal={Acta Crystallogr. Sect. C Cryst. Struct. Commun.},
  volume={52},
  number={6},
  pages={1323--1325},
  year={1996},
  doi={10.1107/S0108270195016738},
  publisher={International Union of Crystallography}
}

@article{Mizoguchi2011,
  title = {{Coexistence of Light and Heavy Carriers Associated with Superconductivity and Antiferromagnetism in ${\mathrm{CeNi}}_{0.8}{\mathrm{Bi}}_{2}$ with a Bi Square Net}},
  author = {Mizoguchi, Hiroshi and Matsuishi, Satoru and Hirano, Masahiro and Tachibana, Makoto and Takayama-Muromachi, Eiji and Kawaji, Hitoshi and Hosono, Hideo},
  journal = {Phys. Rev. Lett.},
  volume = {106},
  issue = {5},
  pages = {057002},
  numpages = {4},
  year = {2011},
  doi={10.1103/PhysRevLett.106.057002},
  month = {Feb},
  publisher = {American Physical Society}
}

@article{seo2012pressure,
  title={{Pressure effects on the heavy-fermion antiferromagnet CeAuSb$_2$}},
  author={Seo, S and Sidorov, VA and Lee, H and Jang, D and Fisk, Z and Thompson, Joe David and Park, T},
  journal={Phys. Rev. B},
  volume={85},
  number={20},
  pages={205145},
  year={2012},
  doi={10.1103/PhysRevB.85.205145},
  publisher={APS}
}

@article{jesus2014evolution,
  title={{Evolution of the magnetic properties along the RCuBi2 (R= Ce, Pr, Nd, Gd, Sm) series of intermetallic compounds}},
  author={Jesus, Camilo Bruno Ramos de and Piva, M{\'a}rio Moda and Rosa, Priscila Ferrari Silveira and Adriano, Cris and Pagliuso, PG},
  journal={J. Appl. Phys.},
  volume={115},
  number={17},
  pages={17E115},
  year={2014},
  doi={10.1063/1.4860657},
  publisher={AIP Publishing}
}

@article{Piva2020,
   author = {M. M. Piva and R. Tartaglia and G. S. Freitas and J. C. Souza and D. S. Christovam and S. M. Thomas and J. B. Leaõ and W. Ratcliff and J. W. Lynn and C. Lane and J. X. Zhu and J. D. Thompson and P. F.S. Rosa and C. Adriano and E. Granado and P. G. Pagliuso},
   doi = {10.1103/PhysRevB.101.214431},
   issn = {24699969},
   issue = {21},
   journal = {Phys. Rev. B},
   month = {6},
   publisher = {American Physical Society},
   title = {Electronic and magnetic properties of stoichiometric {CeAuBi$_2$}},
   volume = {101},
   pages = {214431},
   year = {2020}
}

@article{wang2021topologically,
  title={Topologically nontrivial type-{I} and type-{II} nodal-line states in magnetic configurations of square-net pnictide {CeAuBi$_2$}},
  author={Wang, Athena and Luo, Xuan},
  journal={Comput. Mater. Sci.},
  volume={194},
  pages={110434},
  year={2021},
  doi={10.1016/j.commatsci.2021.110434},
  publisher={Elsevier}
}

@article{mendoncca2008tuning,
  title={Tuning the pressure-induced superconducting phase in doped {CeRhIn$_5$}},
  author={Mendon{\c{c}}a Ferreira, L and Park, T and Sidorov, V and Nicklas, M and Bittar, EM and Lora-Serrano, R and Hering, EN and Ramos, SM and Fontes, MB and Baggio-Saitovich, E and others},
  journal={Phys. Rev. Lett.},
  volume={101},
  number={1},
  pages={017005},
  year={2008},
  doi={10.1103/physrevlett.101.017005},
  publisher={APS}
}

@article{steglich1979superconductivity,
  title={Superconductivity in the presence of strong {Pauli} paramagnetism: {CeCu$_2$Si$_2$}},
  author={Steglich, Frank and Aarts, J and Bredl, CD and Lieke, W and Meschede, D and Franz, W and Sch{\"a}fer, H},
  journal={Phys. Rev. Lett.},
  volume={43},
  number={25},
  pages={1892},
  year={1979},
  doi={10.1103/PhysRevLett.43.1892},
  publisher={APS}
}

@article{Thomas2016,
  title = {{Hall} effect anomaly and low-temperature metamagnetism in the {Kondo} compound {${\mathrm{CeAgBi}}_{2}$}},
  author = {Thomas, S. M. and Rosa, P. F. S. and Lee, S. B. and Parameswaran, S. A. and Fisk, Z. and Xia, J.},
  journal = {Phys. Rev. B},
  volume = {93},
  issue = {7},
  pages = {075149},
  numpages = {8},
  year = {2016},
  month = {Feb},
  doi={10.1103/PhysRevB.93.075149},
  publisher = {American Physical Society}
  }

@article{feringa2022spin,
  title={{Spin-flop transition in the quasi-two-dimensional antiferromagnet MnPS$_3$ detected via thermally generated magnon transport}},
  author={Feringa, Frank and Vink, JM and Van Wees, BJ},
  journal={Phys. Rev. B},
  volume={106},
  number={22},
  pages={224409},
  year={2022},
  doi={10.1103/PhysRevB.106.224409},
  publisher={APS}
}

@article{keffer1953spin,
  title={Spin waves in ferromagnetic and antiferromagnetic materials},
  author={Keffer, F and Kaplan, H and Yafet, Y},
  journal={Am. J. Phys.},
  volume={21},
  number={4},
  pages={250--257},
  year={1953},
  doi={10.1119/1.1933416},
  publisher={American Association of Physics Teachers}
}

@article{li2016possible,
  title={Possible ground states and parallel magnetic-field-driven phase transitions of collinear antiferromagnets},
  author={Li, Hai-Feng},
  journal={npj Comput. Mater.},
  volume={2},
  number={1},
  pages={16032},
  year={2016},
  doi={10.1038/npjcompumats.2016.32},
  publisher={Nature Publishing Group}
}

@article{lipika2026complex,
  title={{Complex spin dynamics induced metamagnetic phase transitions in Dirac semimetal EuAuBi}},
  author={Lipika and Singh, Shobha and Saraswati, Anyesh and Chahar, Vikas and Sun, Yan and Manuel, Pascal and Adroja, Devashibhai and Schnelle, Walter and Kumar, Nitesh and Sannigrahi, Jhuma and others},
  journal={Phys. Rev. B},
  volume={113},
  number={10},
  pages={104406},
  year={2026},
  doi={10.1103/dt26-56xc},
  publisher={APS}
}

@article{chun2001reentrant,
  title={Reentrant spin glass behavior in layered manganite {La$_{1.2}$Sr$_{1.8}$Mn$_2$O$_7$} single crystals},
  author={Chun, SH and Lyanda-Geller, Y and Salamon, MB and Suryanarayanan, R and Dhalenne, G and Revcolevschi, A},
  journal={J. Appl. Phys.},
  volume={90},
  number={12},
  pages={6307--6311},
  year={2001},
  doi={10.1063/1.1419260},
  publisher={American Institute of Physics}
}

@article{kumar2021multiple,
  title={{Multiple magnetic phase transitions with different universality classes in bilayer La$_{1.4}$Sr$_{1.6}$Mn$_2$O$_7$ manganite}},
  author={Kumar, Birendra and Tiwari, Jeetendra Kumar and Chauhan, Harish Chandr and Ghosh, Subhasis},
  journal={Sci. Rep.},
  volume={11},
  number={1},
  pages={21184},
  year={2021},
  doi={10.1038/s41598-021-00544-8},
  publisher={Nature Publishing Group UK London}
}

@article{petrovic2002anisotropic,
  title={Anisotropic properties of rare-earth dibismites},
  author={Petrovic, C and Bud’ko, SL and Canfield, PC},
  journal={J. Magn. Magn. Mater.},
  volume={247},
  number={3},
  pages={270--278},
  year={2002},
  doi={10.1016/S0304-8853(02)00278-0},
  publisher={Elsevier}
}

@article{myers1999systematic,
  title={{Systematic study of anisotropic transport and magnetic properties of RAgSb$_2$ (R= Y, La--Nd, Sm, Gd--Tm)}},
  author={Myers, KD and Bud'Ko, SL and Fisher, IR and Islam, Z and Kleinke, H and Lacerda, AH and Canfield, PC},
  journal={J. Magn. Magn. Mater.},
  volume={205},
  number={1},
  pages={27--52},
  year={1999},
  doi={10.1016/S0304-8853(99)00472-2},
  publisher={Elsevier}
}

@article{petrovic2003anisotropic,
  title={Anisotropic properties of rare earth silver dibismites},
  author={Petrovic, C and Bud’ko, SL and Strand, JD and Canfield, PC},
  journal={J. Magn. Magn. Mater.},
  volume={261},
  number={1-2},
  pages={210--221},
  year={2003},
  doi={10.1016/S0304-8853(02)01476-2},
  publisher={Elsevier}
}

@article{abrikosov1998quantum,
  title={Quantum magnetoresistance},
  author={Abrikosov, AA},
  journal={Phys. Rev. B},
  volume={58},
  number={5},
  pages={2788},
  year={1998},
  doi={10.1103/PhysRevB.58.2788},
  publisher={APS}
}

@article{parish2003non,
  title={Non-saturating magnetoresistance in heavily disordered semiconductors},
  author={Parish, MM and Littlewood, PB},
  journal={Nature},
  volume={426},
  number={6963},
  pages={162--165},
  year={2003},
  doi={10.1038/nature02073},
  publisher={Nature Publishing Group UK London}
}

@article{xiang2021unusual,
  title={{Unusual high-field metal in a Kondo insulator}},
  author={Xiang, Ziji and Chen, Lu and Chen, Kuan-Wen and Tinsman, Colin and Sato, Yuki and Asaba, Tomoya and Lu, Helen and Kasahara, Yuichi and Jaime, Marcelo and Balakirev, Fedor and others},
  journal={Nat. Phys.},
  volume={17},
  number={7},
  pages={788--793},
  year={2021},
  doi={10.1038/s41567-021-01216-0},
  publisher={Nature Publishing Group UK London}
}

@article{dzero2012theory,
  title={{Theory of topological Kondo insulators}},
  author={Dzero, Maxim and Sun, Kai and Coleman, Piers and Galitski, Victor},
  journal={Phys. Rev. B},
  volume={85},
  number={4},
  pages={045130},
  year={2012},
  doi={10.1103/PhysRevB.85.045130},
  publisher={APS}
}

@article{chen2015magnetoresistance,
  title={{Magnetoresistance evidence of a surface state and a field-dependent insulating state in the Kondo insulator SmB$_6$}},
  author={Chen, F and Shang, C and Jin, Z and Zhao, D and Wu, YP and Xiang, ZJ and Xia, ZC and Wang, AF and Luo, XG and Wu, T and others},
  journal={Phys. Rev. B},
  volume={91},
  number={20},
  pages={205133},
  year={2015},
  doi={10.1103/PhysRevB.91.205133},
  publisher={APS}
}

@article{RAUCHSCHWALBE1987347,
title = {Magnetoresistance of Ce-based Kondo lattices: {CeCu$_2$Si$_2$} and {CeAl$_3$}},
journal = {J. Magn. Magn. Mater.},
volume = {63-64},
pages = {347-350},
year = {1987},
issn = {0304-8853},
doi = {https://doi.org/10.1016/0304-8853(87)90607-X},
url = {https://www.sciencedirect.com/science/article/pii/030488538790607X},
author = {U. Rauchschwalbe and F. Steglich and A. {de Visser} and J.J.M. Franse}
}

@article{hundley1993magnetoresistance,
  title={{Magnetoresistance of the Kondo insulator Ce$_3$Bi$_4$Pt$_3$}},
  author={Hundley, Michael Frederic and Lacerda, A and Canfield, PC and Thompson, Joe David and Fisk, Z},
  journal={Physica B: Condens. Matter},
  volume={186},
  pages={425--427},
  year={1993},
  doi={10.1016/0921-4526(93)90593-U},
  publisher={Elsevier}
}

@article{ritzinger2023anisotropic,
  title={Anisotropic magnetoresistance: materials, models and applications},
  author={Ritzinger, Philipp and V{\`y}born{\`y}, Karel},
  journal={R. Soc. Open Sci.},
  volume={10},
  number={10},
  pages={230564},
  year={2023},
  doi={10.1098/rsos.230564},
  publisher={The Royal Society}
}

@article{yang2021colossal,
  title={{Colossal angular magnetoresistance in the antiferromagnetic semiconductor EuTe$_2$}},
  author={Yang, Huali and Liu, Qing and Liao, Zhaoliang and Si, Liang and Jiang, Peiheng and Liu, Xiaolei and Guo, Yanfeng and Yin, Junjie and Wang, Meng and Sheng, Zhigao and others},
  journal={Phys. Rev. B},
  volume={104},
  number={21},
  pages={214419},
  year={2021},
  doi={10.1103/PhysRevB.104.214419},
  publisher={APS}
}

@article{marti2014room,
  title={Room-temperature antiferromagnetic memory resistor},
  author={Marti, X and Fina, I and Frontera, C and Liu, Jian and Wadley, P and He, Qing and Paull, RJ and Clarkson, JD and Kudrnovsk{\`y}, J and Turek, I and others},
  journal={Nat. Mater.},
  volume={13},
  number={4},
  pages={367--374},
  year={2014},
  doi={10.1038/nmat3861},
  publisher={Nature Publishing Group UK London}
}

@article{wang2019giant,
  title={Giant anisotropic magnetoresistance and nonvolatile memory in canted antiferromagnet {Sr$_2$IrO$_4$}},
  author={Wang, Haowen and Lu, Chengliang and Chen, Jun and Liu, Yong and Yuan, SL and Cheong, Sang-Wook and Dong, Shuai and Liu, Jun-Ming},
  journal={Nat. Commun.},
  volume={10},
  number={1},
  pages={2280},
  year={2019},
  doi={10.1038/s41467-019-10299-6},
  publisher={Nature Publishing Group UK London}
}

@article{soh2019magnetic,
  title={{Magnetic and electronic structure of Dirac semimetal candidate EuMnSb$_2$}},
  author={Soh, J-R and Manuel, P and Schr{\"o}ter, NMB and Yi, CJ and Orlandi, F and Shi, YG and Prabhakaran, D and Boothroyd, AT},
  journal={Phys. Rev. B},
  volume={100},
  number={17},
  pages={174406},
  year={2019},
  doi={10.1103/PhysRevB.100.174406},
  publisher={APS}
}

@article{bolte2005magnetotransport,
  title={Magnetotransport through magnetic domain patterns in permalloy rectangles},
  author={Bolte, M and Steiner, M and Pels, C and Barthelmess, M and Kruse, J and Merkt, U and Meier, G and Holz, M and Pfannkuche, D},
  journal={Phys. Rev. B},
  volume={72},
  number={22},
  pages={224436},
  year={2005},
  doi={10.1103/PhysRevB.72.224436},
  publisher={APS}
}

@article{voerman2019origin,
  title={{Origin of the butterfly magnetoresistance in ZrSiS}},
  author={Voerman, JA and Mulder, L and De Boer, JC and Huang, Y and Schoop, LM and Li, Chuan and Brinkman, A},
  journal={Phys. Rev. Mater.},
  volume={3},
  number={8},
  pages={084203},
  year={2019},
  doi={10.1103/PhysRevMaterials.3.084203},
  publisher={APS}
}

@article{feng2006glassy,
  title={Glassy ferromagnetism in {Ni$_3$Sn}-type {Mn$_{3.1}$Sn$_{0.9}$}},
  author={Feng, WJ and Li, D and Ren, WJ and Li, YB and Li, WF and Li, J and Zhang, YQ and Zhang, ZD},
  journal={Phys. Rev. B},
  volume={73},
  number={20},
  pages={205105},
  year={2006},
  doi={10.1103/PhysRevB.73.205105},
  publisher={APS}
}

@article{hiroi2009ferromagnetism,
  title={{Ferromagnetism and spin-glass transitions in the Heusler compounds Ru$_{2-x}$Fe$_x$CrSi}},
  author={Hiroi, Masahiko and Rokkaku, Tsugumi and Matsuda, Kazuhisa and Hisamatsu, Toru and Shigeta, Iduru and Ito, Masakazu and Sakon, Takuo and Koyama, Keiichi and Watanabe, Kazuo and Nakamura, Shintaro and others},
  journal={Phys. Rev. B},
  volume={79},
  number={22},
  pages={224423},
  year={2009},
  doi={10.1103/PhysRevB.79.224423},
  publisher={APS}
}

@book{mydosh1993spin,
  title={Spin glasses: an experimental introduction},
  author={Mydosh, John A},
  year={1993},
  doi={10.1201/9781482295191},
  publisher={CRC press}
}

@article{mulder1981susceptibility,
  title={{Susceptibility of the Cu Mn spin-glass: Frequency and field dependences}},
  author={Mulder, CAM and Van Duyneveldt, AJ and Mydosh, JA},
  journal={Phys. Rev. B},
  volume={23},
  number={3},
  pages={1384},
  year={1981},
  doi={10.1103/PhysRevB.23.1384},
  publisher={APS}
}

@article{giot2008magnetic,
  title={{Magnetic states and spin-glass properties in Bi$_{0.67}$Ca$_{0.33}$MnO$_3$: Macroscopic ac measurements and neutron scattering}},
  author={Giot, Maud and Pautrat, Alain and Andr{\'e}, Gilles and Saurel, Damien and Hervieu, Maryvonne and Rodriguez-Carvajal, Juan},
  journal={Phys. Rev. B},
  volume={77},
  number={13},
  pages={134445},
  year={2008},
  doi={10.1103/PhysRevB.77.134445},
  publisher={APS}
}

@article{manna2011magnetic,
  title={{On the Magnetic Ground State of La$_{0.85}$Sr$_{0.15}$CoO$_3$ Single Crystals}},
  author={Manna, Kaustuv and Samal, Debakanta and Elizabeth, Suja and Bhat, HL and Anil Kumar, PS},
  journal={J. Phys. Chem. C},
  volume={115},
  number={29},
  pages={13985--13990},
  year={2011},
  doi={10.1021/jp201206a},
  publisher={ACS Publications}
}

@article{guo2016spin,
  title={Spin glass behavior in {LaCo$_{1-x}$Rh$_x$O$_3$} (x=0.4, 0.5, and 0.6)},
  author={Guo, Hanjie and Manna, Kaustuv and Luetkens, H and Hoelzel, Markus and Komarek, AC},
  journal={Phys. Rev. B},
  volume={94},
  number={20},
  pages={205128},
  year={2016},
  doi={10.1103/PhysRevB.94.205128},
  publisher={APS}
}

@article{chakrabarty2014cluster,
  title={{Cluster spin glass behavior in geometrically frustrated Zn$_3$V$_3$O$_8$}},
  author={Chakrabarty, Tanmoy and Mahajan, Avinash V and Kundu, Susanta},
  journal={J. Phys.: Condens. Matter},
  volume={26},
  number={40},
  pages={405601},
  year={2014},
  doi={10.1088/0953-8984/26/40/405601},
  publisher={IOP Publishing}
}

@article{maji2011low,
  title={Low temperature cluster glass behavior in {Nd$_5$Ge$_3$}},
  author={Maji, Bibekananda and Suresh, KG and Nigam, AK},
  journal={J. Phys.: Condens. Matter},
  volume={23},
  number={50},
  pages={506002},
  year={2011},
  doi={10.1088/0953-8984/23/50/506002},
  publisher={IOP Publishing}
}

@article{onishi2018deviation,
  title={{Deviation from the Kohler’s rule and Shubnikov--de Haas oscillations in type-II Weyl semimetal WTe$_2$: High magnetic field study up to 56 T}},
  author={Onishi, Shota and Jha, Rajveer and Miyake, Atsushi and Higashinaka, Ryuji and Matsuda, Tatsuma D and Tokunaga, Masashi and Aoki, Yuji},
  journal={AIP Adv.},
  volume={8},
  number={10},
  pages={101330}, 
  year={2018},
  doi={10.1063/1.5043036},
  publisher={AIP Publishing}
}

@article{dasoundhi2024extremely,
  title={{Extremely large magnetoresistance and non-trivial band topology in YSb semimetal}},
  author={Dasoundhi, Mukesh Kumar and Baral, Sonali and Rajput, Indu and Kumar, Devendra and Lakhani, Archana},
  journal={Mater. Today Phys.},
  volume={40},
  pages={101310},
  year={2024},
  doi={10.1016/j.mtphys.2023.101310},
  publisher={Elsevier}
}

@article{PhysRevB.91.224420,
  title = {Structural-modulation-driven spin canting and reentrant glassy magnetic phase in ferromagnetic $\mathrm{L}{\mathrm{u}}_{2}\mathrm{MnNi}{\mathrm{O}}_{6}$},
  author = {Manna, Kaustuv and Bera, A. K. and Jain, Manish and Elizabeth, Suja and Yusuf, S. M. and Anil Kumar, P. S.},
  journal = {Phys. Rev. B},
  volume = {91},
  issue = {22},
  pages = {224420},
  numpages = {7},
  year = {2015},
  month = {Jun},
  publisher = {American Physical Society},
  doi = {10.1103/PhysRevB.91.224420},
  url = {https://link.aps.org/doi/10.1103/PhysRevB.91.224420}
}

@article{manna2014correspondence,
  title={Correspondence between neutron depolarization and higher order magnetic susceptibility to investigate ferromagnetic clusters in phase separated systems},
  author={Manna, Kaustuv and Samal, D and Bera, AK and Elizabeth, Suja and Yusuf, SM and Anil Kumar, PS},
  journal={J. Phys.: Condens. Matter},
  volume={26},
  number={1},
  pages={016002},
  year={2014},
  doi={10.1088/0953-8984/26/1/016002},
  publisher={IOP Publishing}
}

@article{Li2026,
   author = {Zheng Li and Sheng Xu and Yi Yan Wang and Tian Hao Li and Shu Xiang Li and Jin Jin Wang and Jun Jian Mi and Qian Tao and Zhu An Xu},
   doi = {10.1103/6qj9-gv6f},
   issn = {24699969},
   issue = {4},
   journal = {Phys. Rev. B},
   month = {1},
   year = {2026},
   title = {Field-induced magnetic phase transitions and transport anomalies in {GdAlSi}},
   volume = {113},
   pages = {045113},
   publisher = {American Physical Society}
}

@article{Wang2019,
   title = {Magnetic phase diagrams of the ferromagnetic Kondo lattice {CePd$_2$Al$_8$}},
   author = {Le Wang and Cuixiang Wang and Ziyi Liu and Jinguang Cheng and Shanshan Miao and Youting Song and Youguo Shi and Yi Feng Yang},
   journal = prb,
   volume = {100},
   pages = {085122},
   issn = {24699969},
   issue = {8},
   month = {8},
   year = {2019},
   doi = {10.1103/PhysRevB.100.085122},
   publisher = {American Physical Society}
}

@article{Gornicka2026,
   title = {Anisotropic magnetism and Kondo-lattice behavior in the frustrated antiferromagnet {Ce$_3$MgBi$_5$}},
   author = {Karolina Gornicka and Brenden R. Ortiz and Matthew S. Cook and Heda Zhang and Andrew D. Christianson and Andrew F. May},
   issn = {24759953},
   issue = {5},
   journal = {Phys. Rev. Mater.},
   month = {5},
   volume = {10},
   pages = {054413},
   year = {2026},
   doi = {10.1103/g79p-6c1d},
   publisher = {American Physical Society}  
}

@article{Ashtar2026,
   title = {{K$_3$RETe$_2$O$_9$ (RE = Pr, Nd, and Gd–Yb)}: A Family of Rare-Earth Triangular-Lattice Antiferromagnets with Large Lattice Spacings}, 
   author = {Malik Ashtar and Xinyang Liu and Zhaotong Zhuang and Junsen Xiang and Zhaoming Tian and Peijie Sun},
   
   issn = {1520510X},
   issue = {19},
   journal = {Inorg. Chem.},
   month = {5},
   pages = {10474-10483},
   pmid = {42091135},
   doi = {10.1021/acs.inorgchem.6c00076},
   publisher = {American Chemical Society},
   
   volume = {65},
   year = {2026}
}

@article{Wang2010,
   author = {C. H. Wang and J. M. Lawrence and A. D. Christianson and E. A. Goremychkin and V. R. Fanelli and K. Gofryk and E. D. Bauer and F. Ronning and J. D. Thompson and N. R. De Souza and A. I. Kolesnikov and K. C. Littrell},
   doi = {10.1103/PhysRevB.81.235132},
   issn = {10980121},
   issue = {23},
   journal = {Phys. Rev. B},
   month = {6},
   title = {Kondo behavior, ferromagnetic correlations, and crystal fields in the heavy-fermion compounds {Ce$_3$X(X=In,Sn )}},
   volume = {81},
   pages = {235132},
   year = {2010}
}

@article{Chen2017,
   author = {Jian Chen and Zhen Wang and Shiyi Zheng and Chunmu Feng and Jianhui Dai and Zhu'an Xu},
   doi = {10.1038/srep41853},
   issn = {20452322},
   journal = {Sci. Rep.},
   month = {2},
   pmid = {28157184},
   publisher = {Nature Publishing Group},
   title = {Antiferromagnetic Kondo lattice compound {CePt$_3$P}},
   volume = {7},
   pages={41853},
   year = {2017}
}

@article{Hodovanets2026,
   author = {Halyna Hodovanets and Hyunsoo Kim and Tristin Metz and Yasuyuki Nakajima and Christopher J. Eckberg and Kefeng Wang and Jie Yong and Shanta R. Saha and David Graf and Nicholas P. Butch and Thomas Vojta and Johnpierre Paglione},
   doi = {10.1103/1QCC-BY7Y},
   issn = {24699969},
   issue = {5},
   journal = {Phys. Rev. B},
   month = {2},
   publisher = {American Physical Society},
   title = {Magnetic field tuned magnetic order and metamagnetic criticality in nonstoichiometric {CeAuBi$_2$}},
   volume = {113},
   pages = {054432},
   year = {2026}
}

@article{Samwer1976,
   author = {K Samwer and K Winzer},
   doi = {10.1007/BF01420889},
   journal = {Z. Physik B},
   pages = {269-274},
   title = {Magnetoresistivity of the Kondo-System {(La, Ce)B\textsubscript{6}}},
   volume = {25},
   year = {1976}
}

@article{Thompson1986,
   author = {J D Thompson and R D Parks and H Borges},
   doi = {10.1016/0304-8853(86)90627-X},
   journal = {J. Magn. Magn. Mater.},
   pages = {377-378},
   title = {EFFECT OF PRESSURE ON THE NhEL TEMPERATURE OF KONDO-LATTICE SYSTEMS},
   volume = {54},
   year = {1986}
}

@article{Lin1987,
   author = {C L Lin and A Wallash and J E Crow and T Mihalisin and P Schlottmann},
   doi = {10.1103/PhysRevLett.58.1232},
   journal = prl,
   title = {Heavy-Fermion Behavior and the Single-Ion Kondo Model},
   volume = {58},
   pages = {1232--1235},
   year = {1987}
}

@article{Ruvalds1988,
   author = {J Ruvalds' and Q G Sheng},
   doi = {10.1103/PhysRevB.37.1959},
   issue = {4},
   journal = {Phys. Rev. B},
   title = {Magaetoresistance in heavy-fermion alloys},
   volume = {8},
   pages = {1959--1968},
   year = {1988}
}

@article{Sampathkumaran,
   author = {E V Sampathkumaran and Y Nakazawa and M Ishikawa and R Vijayaraghavan},
   doi = {10.1103/PhysRevB.40.11452},
   journal = {Phys. Rev. B},
   title = {Nature of 4f magnetism in {Ce$_{1-x}$La$_x$Pd$_2$Si$_2$}},
   volume = {40},
   pages = {11452(R)--11455(R)},
   year = {1989}
}

@article{Nakatsuji2004,
   author = {Satoru Nakatsuji and David Pines and Zachary Fisk},
   doi = {10.1103/PhysRevLett.92.016401},
   issn = {10797114},
   issue = {1},
   journal = prl,
   pages = {4},
   title = {Two Fluid Description of the Kondo Lattice},
   volume = {92},
   year = {2004}
}

@article{Pikul2012,
  author  = {Pikul, A. P. and Stockert, U. and Steppke, A. and Cichorek, T. and Hartmann, S. and Caroca-Canales, N. and Oeschler, N. and Brando, M. and Geibel, C. and Steglich, F.},
  title   = {Single-ion Kondo scaling of the coherent Fermi liquid regime in {Ce$_{1-x}$La$_x$Ni$_2$Ge$_2$}},
  journal = {Phys. Rev. Lett.},
  volume  = {108},
  pages   = {066405},
  year    = {2012},
  doi     = {10.1103/PhysRevLett.108.066405}
}

@article{Hodovanets2015,
   author = {H. Hodovanets and S. L. Bud'Ko and W. E. Straszheim and V. Taufour and E. D. Mun and H. Kim and R. Flint and P. C. Canfield},
   doi = {10.1103/PhysRevLett.114.236601},
   issn = {10797114},
   issue = {23},
   journal = {Phys. Rev. Lett.},
   month = {6},
   publisher = {American Physical Society},
   title = {Remarkably robust and correlated coherence and antiferromagnetism in {(Ce$_{1-x}$La$_{x}$)Cu$_2$Ge$_2$}},
   volume = {114},
   pages = {236601},
   year = {2015}
}

@article{Hu2008,
   author = {Rongwei Hu and K. J. Thomas and Y. Lee and T. Vogt and E. S. Choi and V. F. Mitrović and R. P. Hermann and F. Grandjean and P. C. Canfield and J. W. Kim and A. I. Goldman and C. Petrovic},
   doi = {10.1103/PhysRevB.77.085212},
   issn = {10980121},
   issue = {8},
   journal = {Phys. Rev. B},
   month = {2},
   title = {Colossal positive magnetoresistance in a doped nearly magnetic semiconductor},
   volume = {77},
   year = {2008},
   pages = {085212},
}

@article{Pavlosiuk2015,
   author = {Orest Pavlosiuk and Dariusz Kaczorowski and Piotr Wi{\"s}niewski},
   doi = {10.1038/srep09158},
   issn = {20452322},
   journal = {Sci. Rep.},
   month = {3},
   publisher = {Nature Publishing Group},
   title = {Shubnikov - De Haas oscillations, weak antilocalization effect and large linear magnetoresistance in the putative topological superconductor {LuPdBi}},
   volume = {5},
   pages={9158},
   year = {2015}
}

\end{document}